\documentclass[aps,prl,preprint,superscriptaddress]{revtex4-2}

\usepackage[T1]{fontenc}
\usepackage{amsmath,amssymb}
\usepackage{graphicx}
\usepackage{hyperref}                
\usepackage{placeins}

\begin{document}

\title{Silent breakdown of the perturbative polarization expansion in the terahertz regime}

\author{Yao Lu}
\email{yaolu@nankai.edu.cn}
\affiliation{Key Laboratory of Weak-Light Nonlinear Photonics, Ministry of Education, School of Physics and TEDA Applied Physics Institute, Nankai University, Tianjin, 300071, China}
\affiliation{Collaborative Innovation Center of Extreme Optics, Shanxi University, Taiyuan, Shanxi, 030006, China}
\author{Jiawei Zhang}
\affiliation{Key Laboratory of Weak-Light Nonlinear Photonics, Ministry of Education, School of Physics and TEDA Applied Physics Institute, Nankai University, Tianjin, 300071, China}

\author{Qiang Wu}
\email{wuqiang@nankai.edu.cn}
\author{Jingjun Xu}
\email{jjxu@nankai.edu.cn}
\affiliation{Key Laboratory of Weak-Light Nonlinear Photonics, Ministry of Education, School of Physics and TEDA Applied Physics Institute, Nankai University, Tianjin, 300071, China}
\affiliation{Collaborative Innovation Center of Extreme Optics, Shanxi University, Taiyuan, Shanxi, 030006, China}

\date{\today}


\begin{abstract}
Nonlinear optics expands the polarization in powers of the drive and retains only the lowest-order terms. This truncation is safe at optical frequencies, where the convergence field lies far above the damage threshold. In the terahertz regime, phonon polaritons enhance the nonlinear susceptibilities by orders of magnitude, lowering the convergence field into the sub-damage range and opening a window in which the expansion fails while the crystal remains recoverable. We term this window silent breakdown. At resonance, the nearest singularities of the driven lattice mode are a complex-conjugate pair of branch points below the multistability threshold, so the response remains smooth and single-valued even though the $\chi^{\left( n \right)}$ description fails. The convergence field scales as $Q^{-3/2}$, so sharpening a resonance enhances the nonlinearity but narrows the range over which the expansion applies. It follows from the mode shift $S$ alone, $E_{\mathrm{conv}}=E_{\mathrm{ref}}(3\sqrt{3}QS)^{-1/2}$, giving 2$\sim$9 MV/cm for $\mathrm{LiNbO}_{3}$. Since resonant enhancement is central to terahertz nonlinear optics, this boundary defines the field range over which susceptibility-based descriptions are quantitatively reliable.
\end{abstract}

\maketitle

\section*{Introduction}
Nonlinear optics rests on expanding the polarization in powers of the driving field and retaining only the lowest orders \cite{boyd2020,butcher1990,orr1971}. At optical frequencies this is safe, since the electronic nonlinearity is weak, and the series converges throughout the sub-damage range. The expansion is nonetheless a power series with a finite radius of convergence that the material can survive beyond. When a resonance enlarges the nonlinear susceptibilities, successive orders become comparable at a lower field, so the convergence field can fall below the damage threshold. In this case, a region absent in conventional optics opens between expansion failure and material damage, as seen in Fig.~\ref{fig:1}(a).

Terahertz (THz) stimulated phonon polaritons (StPhPs) make this region accessible. Resonant THz driving has produced giant nonlinearities in ionic crystals \cite{fang2025,huang2024,lin2022,lu2021,roman2006,sellati2025,subedi2014,yang2025,lu2024}, and drives the lattice into a strongly anharmonic regime \cite{henstridge2022,gubbin2018}, in the extreme far enough that the interatomic potential itself must be reconstructed rather than expanded \cite{hoegen2018}. The nonlinearities are nonetheless reported in the susceptibility language of an expansion that presupposes its own convergence, and the field range in question is already accessible, strong-field THz experiments reaching peak fields well above 1 MV/cm \cite{hirori2011,hoegen2018}. Lowering the resonance from an electronic $(\sim 10^{16}$ s$^{-1})$ to a phonon-polariton mode $(\sim 10^{13}$ s$^{-1})$ increases the driven displacement by 4.5 at fixed field and by about 40 under resonant driving (Supplemental Material~S1A). This displacement sets $\xi$ , the ratio of successive terms in the polarization series, which grows linearly in the cubic channel and quadratically in the quartic one, reversing the ordering familiar from optics. The relevant anharmonicity is mechanical, carried by the lattice potential of Eq.~\eqref{eq:1} rather than by the electronic response \cite{roman2006}, and the delocalized limit, in which the restoring force cancels, does not apply (Supplemental Material~S1B). The perturbative range is left before the structural instability and nonthermal melting studied previously \cite{henstridge2022,feng2025,li2019}, a reversible strongly nonlinear window opens first (Supplemental Material~S1C).

Existing limits on the perturbative regime have been set only by damage or saturation. The ionization criterion \cite{zheltikov2019}, the internal-field bound \cite{khurgin2024}, and control fields computed within the $\chi^{\left( n \right)}$ hierarchy \cite{sellati2025,subedi2014} each identifies a material threshold while assuming that the series itself is valid. The convergence of the polarization series has been examined in electronic, epsilon-near-zero media, where summation to high order gave no evidence of divergence \cite{reshef2017,reshef2019}. The reason is that, in that regime, the driven displacement is small on the lattice scale, so the convergence field derived here lies far above the damage threshold.

In this article, we show that the perturbative expansion in the drive has a finite convergence field set by a single driven anharmonic oscillator. Within the window, the response remains smooth and single-valued, yet departs from every truncated $\chi^{\left( n \right)}$ expansion. We call this silent breakdown, meaning that no discontinuity, hysteresis, or multistability announces the failure, not that the failure is unobservable. The criterion contains no unknown material parameter, and we propose three measurements to test it. Three features distinguish this criterion from the classical Duffing analysis on which it rests: the convergence radius is tied quantitatively to the phonon-polariton enhancement of the susceptibilities through Eq.~\eqref{eq:5}. It is expressed entirely through the measurable mode shift $S$, Eq.~\eqref{eq:17}, with no free lattice parameter; and the failure is silent, a complex branch-point pair at resonance announcing neither jump nor hysteresis.

\section*{Results}
Consider one infrared-active transverse-optical mode with relative sublattice displacement $ x$ , reduced mass $ m$ , effective charge $ Z^{*}$ , and damping $\gamma$ . With the lattice potential

\begin{equation}
V\left( x \right)=\frac{1}{2} m\omega_{\mathrm{TO}}^{2} x^{2}+\frac{1}{3} m\alpha x^{3}+\frac{1}{4} m\beta x^{4}+\ldots,
\label{eq:1}
\end{equation}
whose cubic term is allowed in non-centrosymmetric crystals such as $ \mathrm{LiNbO}_{3}$, but forbidden in centrosymmetric ones. Truncated at quartic order, Eq.~\eqref{eq:1} gives the Duffing oscillator \cite{duffing1918,kovacic2011,nayfeh1979,strogatz2018}, a polynomial potential whose response nevertheless develops the branch points derived below. These are therefore not inherited from the analyticity of Eq.~\eqref{eq:1}. The truncation only bounds the motion at large amplitude and sets the scope of the result. The nonlinear Huang equations \cite{lu2021,huang1951}, whose linear form is the standard polariton theory of ionic crystals \cite{born1954}, are

\begin{equation}
\ddot{x}+\gamma\dot{x}+\omega_{\mathrm{TO}}^{2} x+\alpha x^{2}+\beta x^{3}+\ldots=\frac{Z^{*}}{m} E,
\label{eq:2}
\end{equation}
\begin{equation}
P=NZ^{*} x+\epsilon_{0}\left( \varepsilon_{\infty}-1 \right) E.
\label{eq:3}
\end{equation}
Solving Eq.~\eqref{eq:2} order by order and inserting into Eq.~\eqref{eq:3} generates the $\chi^{\left( n \right)}$ hierarchy. The linear amplitude at drive frequency $\omega$ is

\begin{equation}
x_{0}=\frac{\left( \frac{Z^{*}}{m} \right) E}{\left| \omega_{\mathrm{TO}}^{2}-\omega^{2}-i\gamma\omega \right|},
\label{eq:4}
\end{equation}
whose denominator differs from the Faust-Henry one only by a factor of $\omega_{\mathrm{TO}}^{2}$ \cite{faust1966}. Define $\xi$ , the ratio of anharmonic to harmonic restoring force at amplitude $ x_{0}$ ,

\begin{equation}
\xi\equiv\frac{\left| F_{\mathrm{anh}} \right|}{\left| F_{\mathrm{harm}} \right|}=\begin{cases}\frac{\alpha x_{0}}{\omega_{\mathrm{TO}}^{2}}\equiv\xi_{3} & \text{cubic (noncentrosymmetric)} \\ \frac{\beta x_{0}^{2}}{\omega_{\mathrm{TO}}^{2}}\equiv\xi_{4} & \text{quartic (centrosymmetric)}\end{cases}
\label{eq:5}
\end{equation}

Each perturbative order contributes one anharmonic vertex and one power of the driven displacement, so successive terms are controlled by $\xi$ . Asymptotically, and up to a constant fixed by the singularity structure below,

\begin{equation}
\frac{\left| P^{\left( n+1 \right)} \right|}{\left| P^{\left( n \right)} \right|}\sim\frac{\xi}{\xi_{c}}[1+O(1/n) ],
\label{eq:6}
\end{equation}
where $\xi_{c}$ is given in Eq.~\eqref{eq:10} and the $ O(1/n)$ correction follows from the transfer theorem for algebraic singularities \cite{flajolet2009,bender1978} (Supplemental Material~S2). Thus $\xi$ is the effective expansion parameter, with critical value of $ 0.513/Q$ at resonance. At $ Q\approx 9$ the series already fails near $\xi\approx 6\times{10}^{-2}$ . Setting $\xi=1$ defines a critical displacement $ x_{c}$ ,

\begin{equation}
E\left( \xi=1 \right)=\frac{mx_{c}}{Z^{*}}\left| \omega_{\mathrm{TO}}^{2}-\omega^{2}-i\gamma\omega \right|.
\label{eq:7}
\end{equation}

Eq.~\eqref{eq:7} is a useful scale but is not the convergence radius, because term-by-term magnitude is not analyticity. Eq.~\eqref{eq:6} is Miller's rule in resonance-enhanced, quantitative form \cite{boyd2020,miller1964}. Miller's rule compares successive terms while remaining silent about convergence, whereas in Eq.~\eqref{eq:6} the scale sets the radius. That radius belongs to the exact response, not to any truncation. The series in $ E$ converges up to the nearest singularity of the steady-state amplitude, so a closed amplitude relation, non-perturbative in $ E$ , is required. The minimal form is the single-harmonic version of Eq.~\eqref{eq:2}, with applicability verified by Methods A1. Near the fold, the motion stays periodic and fundamental-dominated, and the nonresonant harmonics renormalize the frequency shift only at relative $ O(\xi)$ , a few-percent correction at $\xi_{c}$ (Methods A2). The quoted range is therefore not an artifact of truncating Eq.~\eqref{eq:1} at quartic order, but is anchored through the interval $ x_{c}=0.5\sim 2$ $\text{\AA}$ (Supplemental Material~S3). The approximation keeps the dominant resonant effect, a drive-induced shift of the mode frequency:

\begin{equation}
x_{0}^{2}\left[ {\left( \Omega\left( x_{0} \right)-\omega^{2} \right)}^{2}+\gamma^{2}\omega^{2} \right]={\left( \frac{Z^{*}}{m} E \right)}^{2},  \Omega\left( x_{0} \right)=\begin{cases}\omega_{\mathrm{TO}}^{2}+\frac{3}{4}\beta x_{0}^{2}, & \text{quartic} \\ \omega_{\mathrm{TO}}^{2}-\frac{5}{6}{\left( \frac{\alpha}{\omega_{\mathrm{TO}}} \right)}^{2} x_{0}^{2}, & \text{cubic}\end{cases}
\label{eq:8}
\end{equation}

Because the anharmonicity enters only as a real shift of the resonance, Eq.~\eqref{eq:8} is an algebraic relation between $ x_{0}$ and $ E$ , and it ceases to be single-valued where $ dE/dx_{0}=0$ . That point is a fold (saddle-node), the jump mechanism of Duffing hysteresis, whose onset obeys Eq.~\eqref{eq:11} \cite{kovacic2011,nayfeh1979,landau1976}. Within the single-harmonic amplitude Eq.~\eqref{eq:8}, the fold is the only singularity available, so it sets the convergence radius. Both ingredients are classical: the fold is the Duffing saddle-node \cite{kovacic2011,nayfeh1979,strogatz2018}, and the radius of convergence of a power series is set by its nearest singularity in the complex drive plane \cite{bender1978,flajolet2009}. The consequence drawn below is not: this singularity is reached at resonance before any jump is possible, and its location is fixed by $ Q$ and the mode shift alone. Written with $ u=x_{0}^{2}$ , $ F=\frac{Z^{*}}{m} E$ , $ G=\gamma\omega$ , $ s=\omega^{2}-\omega_{\mathrm{TO}}^{2}$ , and $ a=\frac{3}{4}\beta$ or $\frac{5}{6}{\left( \frac{\alpha}{\omega_{\mathrm{TO}}} \right)}^{2}$ , Eq.~\eqref{eq:8} is a cubic in $ u$ , $ a^{2} u^{3}\mp 2asu^{2}+\left( s^{2}+G^{2} \right) u-F^{2}=0$ , whose branch points in complex $ F^{2}$ occur where it and its $ u$ -derivative vanish together. Eliminating $ u$ gives the pair (Methods A3)

\begin{equation}
u_{\pm}=\frac{\sigma 2s\pm\sqrt{s^{2}-3G^{2}}}{3a},  \sigma=\begin{cases}+1, & \text{quartic} \\ -1, & \text{cubic}\end{cases}
\label{eq:9}
\end{equation}
and the radius is the value reached by continuing $ u=0$ at $ F^{2}=0$ along the physical branch.

\begin{figure}[!t]
\centering
\includegraphics[width=0.95\columnwidth]{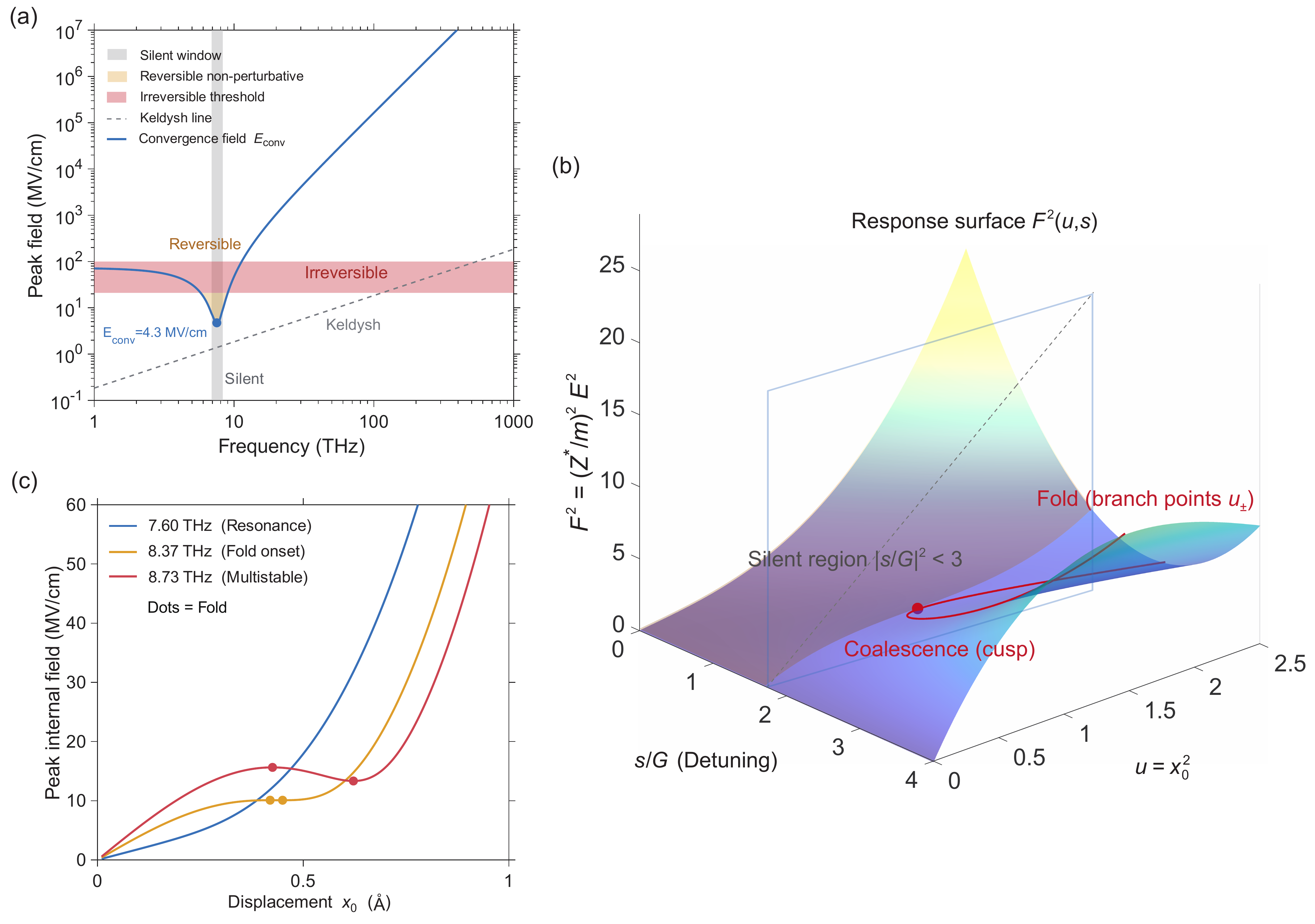}
\caption{Phase diagram and geometry of the convergence boundary. (a) Peak field versus drive frequency for the $ \mathrm{LiNbO}_{3}$ $A_{1}$ polariton mode ($ x_{c}=1$ $\text{\AA}).$ Blue curves: convergence field $ E_{\mathrm{conv}}$ (Eq.~\eqref{eq:9}); red band: irreversible-change threshold; orange region: reversible non-perturbative window; grey dashed line: Keldysh line $\gamma_{K}=1$ . Shaded strip: silent window $|s|<\sqrt{3} G$ . (b) Dimensionless response surface $ F^{2}$ versus detuning $ s/G$ and amplitude $ u$ (Eq.~\eqref{eq:9}). Red curve: fold; shaded slab: silent region $|s/G| <\sqrt{3}$ . (c) Peak internal field versus displacement at 7.60 THz (resonance), 8.37 THz (fold onset) and 8.73 THz; dots mark the fold $\partial E/\partial x_{0}=0$ .\label{fig:1}}
\end{figure}

\begin{figure}[!t]
\centering
\includegraphics[width=0.95\columnwidth]{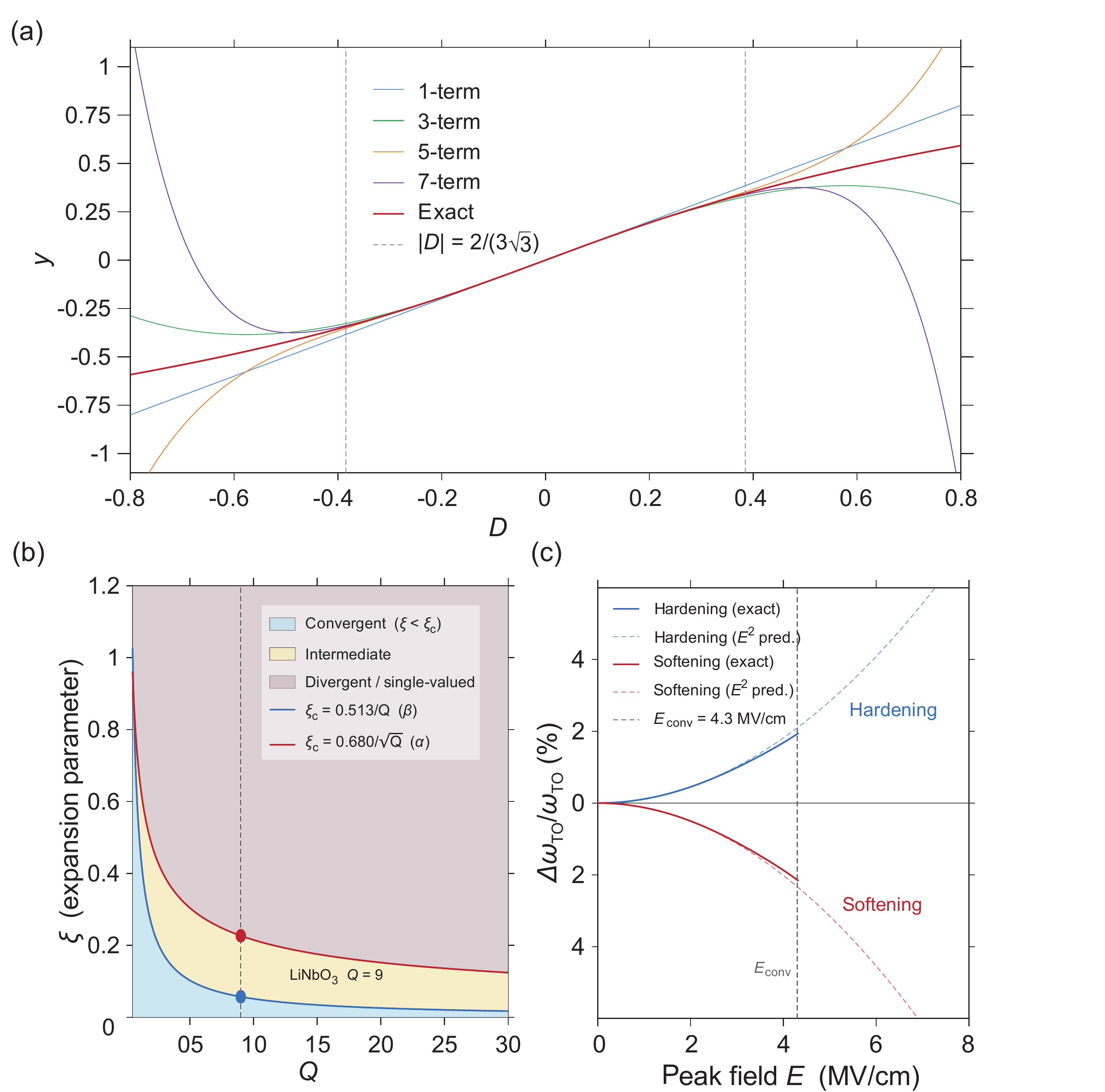}
\caption{Silent failure, critical drive, and resonant mode shift. (a) Exact resonant response $ D=y^{3}+y$ (red curve) against 1-, 3-, 5-, and 7-term truncations (blue, green, orange, purple); vertical dashed lines at $\left| D \right|=\frac{2}{3\sqrt{3}}=0.385$ . Every truncation departs from the smooth, monotone exact solution at this boundary, while the response itself remains single-valued, the silent failure. (b) Critical drive in the $(Q,\xi)$ plane: convergent below the $\beta$ -dominated limit $\xi_{c}=0.513/Q$ , divergent but single-valued above; the $\alpha$ -dominated limit $\xi_{c}=0.680/\sqrt{Q}$ is shown for comparison. The $ \mathrm{LiNbO}_{3}$ polariton mode ($ Q\approx 9$) is marked. (c) Fractional mode shift $\left| \Delta\omega_{\mathrm{TO}}/\omega_{\mathrm{TO}} \right|$ versus peak field for the hardening (upper) and softening (lower) branches. Straight lines are the $ E^{2}$ prediction, and the curvature is the departure from it as the branch point is approached (Methods A4).\label{fig:2}}
\end{figure}

$\xi$ is built on the linear amplitude of Eq.~\eqref{eq:4}, so its critical value at the edge of convergence is $\xi_{c}=\left| a \right|x_{0,\mathrm{lin}}^{2}/\omega_{\mathrm{TO}}^{2}$ , evaluated at $ u_{\mathrm{lin}}=R/{\left| \omega_{\mathrm{TO}}^{2}-\omega^{2}-i\gamma\omega \right|}^{2}$ , rather than at the branch-point amplitude $ u_{\pm}$ . The two differ by a factor of 2/3 at resonance (Methods A3). Three regimes follow (Fig.~\ref{fig:2}(b)): (i) For $\left| s \right|<\sqrt{3} G$ , including exact resonance, the roots of Eq.~\eqref{eq:9} form a complex-conjugate pair, and the branch points lie at complex $ F^{2}$ and there is no jump. At $ s = 0$ , both channels give $ u = \pm iG/(\sqrt{3} a)$ and $ \left| F^{2}\right|= 2G^{3}/(3\sqrt{3} a)$ , i.e.

\begin{equation}
\xi_{c}=\frac{8}{9\sqrt{3}}\frac{1}{Q}=\frac{0.513}{Q} \left( \text{quartic} \right),     \xi_{c}={\left( \frac{4}{5\sqrt{3} Q} \right)}^{\frac{1}{2}}=\frac{0.680}{\sqrt{Q}} \left( \text{cubic} \right),
\label{eq:10}
\end{equation}
(ii) At $\left| s \right|=\sqrt{3} G$ , the pair coalesces onto the real axis, the Duffing saddle-node. $\left| F^{2} \right|$ and $\left| \omega_{\mathrm{TO}}^{2}-\omega^{2}-i\gamma\omega \right|$ both grow fourfold between $ s = 0$ and $ s =\sqrt{3} G$ , so $\xi_{c}$ is unchanged, but it is reached at twice the drive field. At the detuning where a jump first becomes possible, the series has already failed at half that field. (iii) For $\left| s \right|>\sqrt{3} G$ , two real branch points exist with multistability between them, and the expansion is bounded by whichever the physical branch reaches first. Hence at resonance the series fails before any jump occurs, the saddle-node lying outside the radius of convergence, so the breakdown is silent.

These boundaries can be placed in the plane of drive frequency and field [Fig.~\ref{fig:1}(a)]. Evaluating Eq.~\eqref{eq:9} with the $ \mathrm{LiNbO}_{3}$ parameters of Supplemental Material~S3 gives a curve with a minimum of 4.3 MV/cm at 7.6 THz, reproduced by direct time-domain integration to better than 1\% (Methods A1), rising steeply on either side, whereas the threshold for irreversible change is comparatively flat, of order 10$\sim$100 MV/cm. The ordering familiar from optics is inverted inside a band around $\omega_{\mathrm{TO}}$ . The reversible nonperturbative window 5.8$\sim$8.8 THz, within which the silent sub-band $|s|<\sqrt{3}G$ (6.9$\sim$8.3 THz) is the region where no hysteresis or multistability can occur and the breakdown is truly unannounced,

\begin{equation}
\left| \omega-\omega_{\mathrm{TO}} \right|<\frac{\sqrt{3}}{2}\gamma.
\label{eq:11}
\end{equation}
The Keldysh line does not bound the window either. With $\Delta=4$ eV and $ m^{*}= 0.2 m_{e}$ , $\gamma_{K}=1$ falls at only 1.4 MV/cm \cite{keldysh1965}, so across the window $\gamma_{K}=$ 0.2$\sim$0.7 and one is formally inside the tunnelling regime. Yet the tunnelling action falls only from $\sim 110$ to $\sim 27$ and would reach unity only near 200 MV/cm, above the damage threshold. Ionization thus remains exponentially frozen, and the window is one of bound lattice motion (Supplemental Material~S4A).

The convergence field is the magnitude $|F|$ at the branch point reached by continuation from $ u=0$ on the physical sheet, not simply the branch point of smallest modulus (Methods A3). Away from resonance, the physical branch first meets the nearer-in- $ u$ critical point, whose $|F|$ grows as ${\left| s \right|}^{3/2}$ and carries no damping. Only within $\left| s \right|\leq\sqrt{3} G$ , where the branch point stays on the $ G$ -controlled scale $ R\propto G^{3}/a$ , does $ E_{\mathrm{conv}}$ inherit $ Q^{-3/2}$ , with $ G =\gamma\omega_{\mathrm{TO}}$ and $ Q=\frac{\omega_{\mathrm{TO}}}{\gamma}$ ,

\begin{equation}
E_{\mathrm{conv}}\propto\begin{cases} Q^{-\frac{3}{2}} & \text{within}  \left| s \right|\leq\sqrt{3} G, \\ Q^{0} & \text{far off resonance, on either side of it.}\end{cases}
\label{eq:12}
\end{equation}
Raising $ Q$ therefore deepens only the resonant dip and leaves the wings untouched. Resonant enhancement and perturbative validity are one trade-off rather than independent design axes. The same $ Q^{-3/2}$ scaling is consistent with first-principles estimates for monolayer hBN \cite{iyikanat2021} (Supplemental Material~S4B).

Real polar crystals have several coupled modes \cite{kim2025}, and the coupling renormalizes the anharmonicity that sets the bound. For a partner mode $ q$ coupled through $ V_{c}=\lambda X^{2} q$ , the structure invoked in nonlinear phononics \cite{subedi2014,forst2011}, eliminating $ q$ to leading order gives an effective quartic coefficient,

\begin{equation}
\beta_{\mathrm{eff}}=\beta_{0}-\frac{4}{3}\lambda^{2}\left[ \frac{1}{\omega_{R}^{2}}+\frac{1}{2\left( \omega_{R}^{2}-4\omega^{2} \right)} \right],
\label{eq:13}
\end{equation}
with $\omega_{R}$ the partner-mode frequency. Since $ x_{c}=\omega_{\mathrm{TO}}/\sqrt{\beta}$ and $ E_{\mathrm{conv}}\propto{\left| \beta \right|}^{-1/2}$ ,

\begin{equation}
E_{\mathrm{conv}}^{\mathrm{multi}}=E_{\mathrm{conv}}^{\mathrm{single}}\times{\left| \frac{\beta_{0}}{\beta_{\mathrm{eff}}} \right|}^{\frac{1}{2}}.
\label{eq:14}
\end{equation}
The bracket in Eq.~\eqref{eq:13} has a pole at $\omega_{R}=2\omega$ and a zero at $\omega_{R}=\sqrt{8/3}\omega$ , and must be read in magnitude as well as sign. A partner mode inside the window $\sqrt{8/3}\omega<\omega_{R}<2\omega$ lowers $ E_{\mathrm{conv}}$ , most strongly as $\omega_{R}\to 2\omega$ , whereas one outside raises it. In the $\mathrm{LiNbO}_{3}$ spectrum no $A_{1}$-symmetry partner mode lies in the suppression window; the nearest higher infrared-active mode ($\approx$13 THz) is of $E$ symmetry and is forbidden by symmetry from coupling through Eq.~\eqref{eq:13} ($A_{1}\otimes E=E\neq A_{1}$). Every $A_1$-coupled infrared-active mode lowers $\left| \beta_{\mathrm{eff}} \right|$, so the single-mode values of Table~\ref{tab:1} are conservative lower bounds (Methods A5, Supplemental Material~S3).

For $\xi$ below $\xi_{c}$ , the leading signatures are analytic in $ E^{2}$ . Harmonic balance on Eq.~\eqref{eq:2} gives, for the quartic channel,

\begin{equation}
\frac{\Delta\omega_{\mathrm{TO}}}{\omega_{\mathrm{TO}}}=+\frac{3}{8}\xi_{4},
\label{eq:15}
\end{equation}
i.e. hardening, and for the cubic channel, through the rectified displacement,

\begin{equation}
\frac{\Delta\omega_{\mathrm{TO}}}{\omega_{\mathrm{TO}}}=-\frac{5}{12}\xi_{3}^{2},
\label{eq:16}
\end{equation}
i.e. softening (Fig.~\ref{fig:2}(c)). Because $\xi_{3}\propto x_{0}$ and $\xi_{4}\propto x_{0}^{2}$ , both channels give $\Delta\omega_{\mathrm{TO}}/\omega_{\mathrm{TO}}\propto E^{2}$ , while their relative magnitude is a field-independent material constant. No sign reversal is predicted at this order, so the sign is falsifiable. For a non-centrosymmetric crystal, both $\alpha$ and $\beta$ enter the steady state only through $ a_{\mathrm{eff}}=3\beta/4-5\alpha^{2}/(6\omega_{\mathrm{TO}}^{2})$ , and it is $ \mathrm{sign}(a_{\mathrm{eff}})$ that the measured sign of the shift determines. Since $ x_{0}\propto E$ , measuring the fractional shift $ S\equiv|\Delta\omega_{\mathrm{TO}}/\omega_{\mathrm{TO}}|$ at a reference field $ E_{\mathrm{ref}}$ fixes $\xi_{\mathrm{ref}}$ , and requiring $\xi(E_{\mathrm{conv}}) =\xi_{c}$ gives the single expression

\begin{equation}
E_{\mathrm{conv}}=E_{\mathrm{ref}}\sqrt{\frac{1}{3\sqrt{3} QS}}
\label{eq:17}
\end{equation}

This is identical for both channels and independent of which anharmonic term dominates. The bound is set by $ Q$ and by the shift at a single reference field alone, and at resonance it is exact within the model of Eqs.~\eqref{eq:1}--\eqref{eq:3}. Read with $ S$ held fixed, Eq.~\eqref{eq:17} gives $ E_{\mathrm{conv}}\propto Q^{-1/2}$ . The $ Q^{-3/2}$ of Eq.~\eqref{eq:12} is recovered by letting $ S$ take its resonant value, since at fixed $ E_{\mathrm{ref}}$ the linear amplitude grows as $ x_{0}\propto Q$ and hence $ S\propto Q^{2}$ . Table~\ref{tab:1} evaluates Eq.~\eqref{eq:17} for the $ \mathrm{LiNbO}_{3}$ polariton mode as a function of $S$ at $ E_{\mathrm{ref}}=1$ MV/cm. The shift has not yet been measured for this mode; it is at present inferred from the anchoring interval $ x_{c}=0.5\sim2$ $\text{\AA}$, which restricts it to $ S\approx 2.8\times{10}^{-4}\sim 4.5\times{10}^{-3}$ and hence gives $ E_{\mathrm{conv}}\approx 2.2\sim 9.0$ MV/cm (Supplemental Material~S3). This spread is a systematic band rather than a statistical error, and it is the width of the anchoring interval propagated through $ E_{\mathrm{conv}}\propto S^{-1/2}$. It collapses to a single value as soon as $ S$ is measured, which is the purpose of measurement (1) below. No lattice parameter enters Eq.~\eqref{eq:17} once $ S$ is known. Three caveats act in different directions and are quantified in Supplemental Material~S3.

\begin{table}[tb]
\centering
\caption{Convergence field from Eq.~\eqref{eq:17} for the $ \mathrm{LiNbO}_{3}$ $A_{1}$ polariton mode, as a function of the fractional mode shift $ S$ at $ E_{\mathrm{ref}}=1$ MV/cm. Values are the single-mode convergence fields in the monochromatic, constant-$\gamma$ model, independent of which anharmonic channel dominates. The range 2.2$\sim$9.0 MV/cm quoted in the text is the sub-range selected by the present anchoring interval $ S=2.8\times10^{-4}\sim4.5\times10^{-3}$; a direct measurement of $ S$ selects one row. Multimode coupling shifts them as discussed above and in Supplemental Material~S3.\label{tab:1}}
\begin{tabular}{lcccc}
\hline
$S$ at 1 MV/cm & 	$10^{-4}$	  & 		$10^{-3}$ 	 & 		$10^{-2}$  & 		$3\times10^{-2}$  \\
(MV/cm) & 15 & 4.7 & 1.5 & 0.84 \\
\hline
\end{tabular}
\end{table}

The displacement is minute, but the readout can be amplified. Just outside the Reststrahlen edge the sensitivity of $\varepsilon(\omega)$ saturates just below 2$Q$,

\begin{equation}
\left| \frac{\partial\mathrm{ln}\varepsilon}{\partial\mathrm{ln}\omega_{\mathrm{TO}}} \right|=\frac{2\omega_{\mathrm{TO}}^{2}\left( \omega_{\mathrm{LO}}^{2}-\omega_{\mathrm{TO}}^{2} \right)}{\left| \omega_{\mathrm{TO}}^{2}-\omega^{2}-i\gamma\omega \right|\left| \omega_{\mathrm{LO}}^{2}-\omega^{2}-i\gamma\omega \right|}\leq 2Q,
\label{eq:18}
\end{equation}
Equation~\eqref{eq:18} is taken at fixed Born charge, so that the longitudinal pole moves with the transverse one and $ \omega_{\mathrm{LO}}^{2}-\omega_{\mathrm{TO}}^{2}$ stays invariant. Holding $ \omega_{\mathrm{LO}}$ fixed instead returns the compact form $ 2\omega_{\mathrm{TO}}^{2}/\left| \omega_{\mathrm{TO}}^{2}-\omega^{2}-i\gamma\omega \right|$, which agrees with Eq.~\eqref{eq:18} to 0.5\% at the transverse frequency but overstates the sensitivity by close to a factor of two far below the edge (Supplemental Material~S5A). Damping, rather than detuning, sets the ceiling. The frequency maximum of Eq.~\eqref{eq:18} sits at $ \omega=1.002\,\omega_{\mathrm{TO}}$ and lies less than 0.1\% above the value at the transverse frequency, so the optimum falls within a linewidth of $ \omega_{\mathrm{TO}}$ (Fig.~S1). For $ \mathrm{LiNbO}_{3}$ ($ \omega_{\mathrm{LO}}/\omega_{\mathrm{TO}}=1.46$, $ Q\approx 9$), Eq.~\eqref{eq:18} gives 17.9 at the transverse frequency, that is, $ 2Q$ to within 0.5\%, enough to render $ x_{0}\sim 10^{-2}$ $\text{\AA}$ measurable (Supplemental Material~S5A). Two consequences follow from the low $ Q$ . Resolving dressed-state sidebands requires a splitting exceeding the linewidth, but Eq.~\eqref{eq:15} at $\xi_{c}$ gives a pull of only $\left( 3/8 \right)\xi_{c}\omega_{\mathrm{TO}}\approx 0.19\gamma$ , so the shift, not the splitting, is the observable of choice. Subharmonics are not expected at the fields of Table~\ref{tab:1}, since the threshold drive for parametric subharmonic instability grows with damping \cite{landau1976}. Sources already reach and exceed 1 MV/cm \cite{hirori2011}.

Three measurements would test the picture:

{\bf{(1) Retrieval of $\xi$ from the mode shift.}} Measure the $ \mathrm{LiNbO}_{3}$ polariton-mode frequency against peak THz field by THz-TDS or time-resolved Raman \cite{hoegen2018,dastrup2017,katayama2012}, confirming the $ E^{2}$ scaling, fix the sign, hence which channel dominates, and extrapolate to the branch point, tuning the probe near the Reststrahlen edge to exploit Eq.~\eqref{eq:18}.

{\bf{(2) Test the susceptibility roll-off.}} In the waveguide of Ref. \cite{lu2021} or the microcavity of Ref. \cite{huang2024}, scan the input power over two to three decades and test whether $\chi_{\mathrm{eff}}^{(2)}$ and $\chi_{\mathrm{eff}}^{(3)}$ stay constant. Because at resonance the anharmonic correction is in quadrature with the linear response, the amplitude departs at $ O(E^{4})$ while the phase departs at $ O(Q\xi)$ . Phase-resolved detection therefore would see a departure $\propto(E/E_{\mathrm{conv}}) ^{2}$ , and amplitude detection one $\propto(E/E_{\mathrm{conv}}) ^{4}$ (Supplemental Material~S5B).

{\bf{(3) Locate the hysteresis with detuning.}} Since Eq.~\eqref{eq:11} forbids a jump at resonance, sweep the frequency at fixed detuning exceeding $(\sqrt{3}/2)\gamma$ , on the side selected by the sign of the shift. The required fractional stability is $\Delta\nu/\nu\sim {10}^{-1}$ over the 5.8$\sim$8.8 THz window and a $\sim 10\%$ field calibration suffices, since $ E_{\mathrm{conv}}$ enters Eq.~\eqref{eq:17} linearly in $ E_{\mathrm{ref}}$ .

In each measurement thermal and coherent nonlinearities are separated by fixing fluence and varying repetition rate (Supplemental Material~S6A), and cascaded second-order contributions are isolated by varying $\Delta k$ (Supplemental Material~S6B). Two boundaries of validity should be stated. We treat a single anharmonic polariton mode in the steady monochromatic state, with $\gamma$ constant and $ V$ truncated at quartic order. At $\xi=\xi_{c}$ the displacement reaches a few tenths of an angstrom, and for $ \mathrm{LiNbO}_{3}$ the shift stays within the anchored interval $ x_{c}=0.5\sim 2$ $\text{\AA}$, so the quoted range is not an artifact of truncation. Several experiments \cite{hoegen2018,handa2024,katayama2012} use single- or few-cycle pulses, whereas the steady-state treatment requires a pulse longer than the polariton coherence time. For $ \mathrm{LiNbO}_{3}$, $\gamma/2\pi=0.84$ THz demands $\gtrsim 0.2$ ps, and under a broadband pulse the breakdown is pushed to a higher peak field. A second, independent failure channel exists whenever a $k$-photon resonance is met, which can tighten the bound even where the fundamental response converges (Supplemental Material~S7). How this criterion relates to previous limits on the perturbative regime is set out in Supplemental Material~S8. 

\section*{Conclusion}
In summary, the $\chi^{(n)}$ description of StPhP-mediated THz nonlinearity has a finite radius of convergence, set by the singularity structure of the driven anharmonic mode rather than by the point where the anharmonic force equals the harmonic one. At resonance the nearest singularity is a complex branch-point pair, so the breakdown is silent, a failure of the expansion rather than of the material. Because the convergence field decreases with increasing $ Q$ , resonant enhancement and perturbative validity are one trade-off, and the bound can be expressed through measurable quantities alone. The argument is not specific to phonon polaritons: any resonantly enhanced nonlinear system whose response is that of a single driven damped oscillator has a convergence radius set by the same branch-point arithmetic, though it must be re-derived for resonances that are not a well-defined oscillator. What is at stake is not how large the THz nonlinearity can be, but whether the susceptibility language applies at all in this regime.

\appendix
\section*{Methods}

\subsection*{A1. Numerical validation of the convergence radius}

The convergence field of Eqs.~\eqref{eq:9} and \eqref{eq:17} is derived from the single-harmonic amplitude relation~\eqref{eq:8}, an approximation that keeps only the fundamental of the motion. To verify that dropping the nonresonant harmonics does not shift the branch point, we integrate the \emph{full} equation of motion, Eq.~\eqref{eq:2}, directly and compare it with Eq.~\eqref{eq:8}.

\begin{figure}[!t]
\centering
\includegraphics[width=0.95\columnwidth]{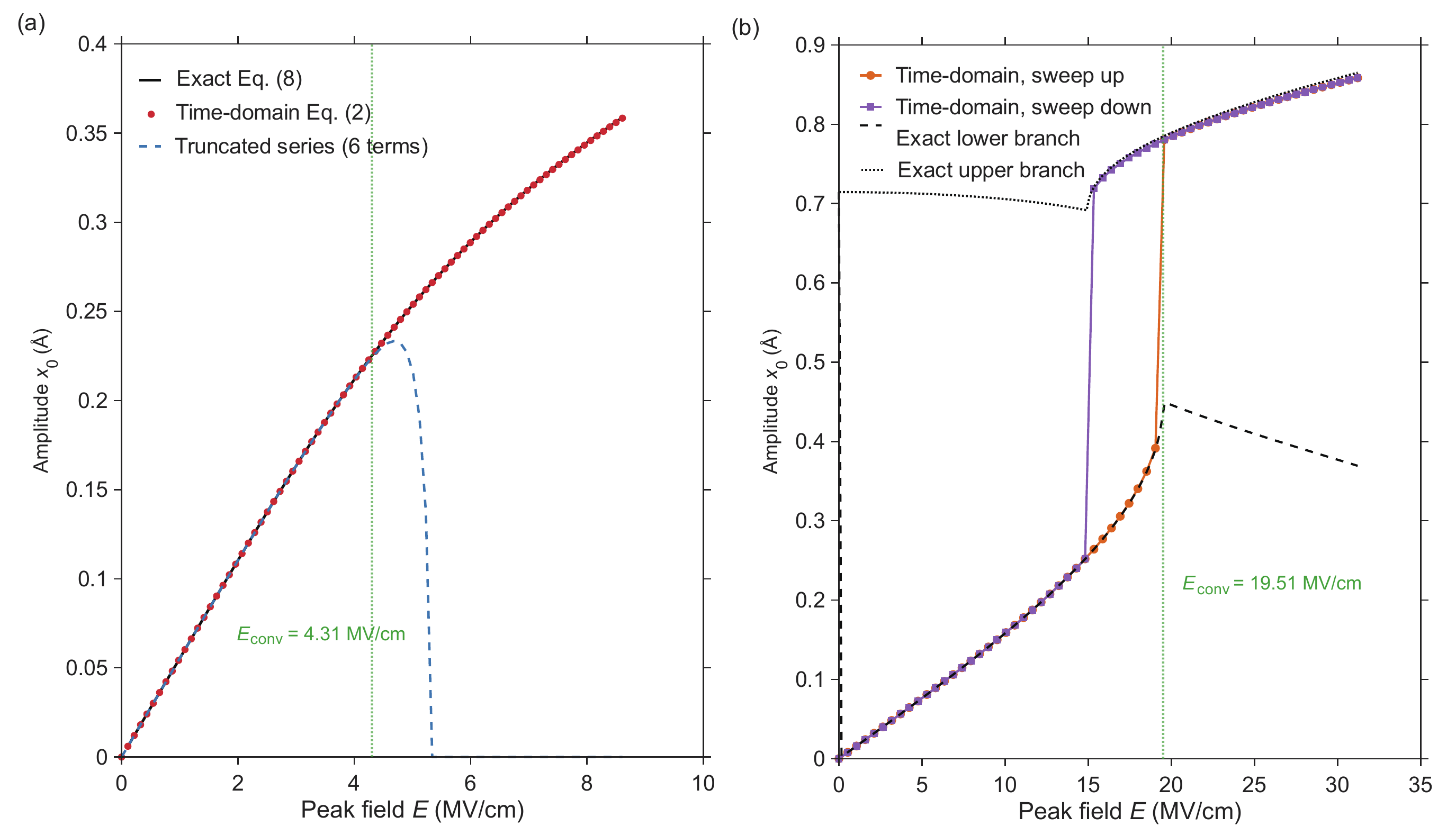}
\caption{Numerical validation of the convergence radius. (a) Steady-state amplitude $|x_{0}|$ versus peak field at exact resonance ($\omega=\omega_{\mathrm{TO}}$). Black: single-harmonic solution, Eq.~\eqref{eq:8}; red dots: direct time-domain integration of the full Eq.~\eqref{eq:2}; blue dashed: six-term truncated expansion in the variable of Fig.~2(a). The two independent solutions coincide, and the truncated series diverges from the smooth response at the convergence field $E_{\mathrm{conv}}=4.3$~MV/cm (green line), the silent breakdown. (b) Off resonance, $\omega_{d}=8.94$~THz ($s=2\sqrt{3}G$): the branch point is real, so the up (orange) and down (purple) sweeps form a hysteresis loop whose fold lies at $E_{\mathrm{conv}}=19.5$~MV/cm (green line), confirming the saddle-node prediction of Eq.~\eqref{eq:9}.\label{fig:3}}
\end{figure}

For the LiNbO$_3$ parameters of Supplemental Material~S3 (Table~S1) the steady state is integrated in the time domain with a fourth-order Runge-Kutta scheme, $\ddot{x}+\gamma\dot{x}+\omega_{\mathrm{TO}}^{2}x+\beta x^{3}=(Z^{\ast}/m)E_{0}\cos\omega t$, starting each field from the previously converged state (state continuation) and measuring the fundamental amplitude over the final fraction of the integration window. The resulting $|x_{0}(E)|$ coincides with the single-harmonic solution of Eq.~\eqref{eq:8} to better than the linewidth (Fig.~\ref{fig:3}a): the single-harmonic truncation is exact for the amplitude at the fields of interest, so the few-percent correction quoted in the main text is a conservative upper bound. The same time-domain solution exposes the breakdown in physical units. The six-term truncated expansion in the variable of Fig.~2(a) departs from the smooth response at the field where the branch-point pair is reached, while the exact response stays single-valued, the silent failure of Fig.~2(a).

Repeating the integration off resonance, inside the multistability window $|s|>\sqrt{3}G$ of Eq.~\eqref{eq:11}, makes the saddle-node explicit: the up- and
down-field sweeps form a hysteresis loop whose fold sits at the field predicted by Eq.~\eqref{eq:9} (Fig.~\ref{fig:3}b). At exact resonance the fold is absent and the branch points are complex, so no loop forms and the breakdown is silent, in agreement with the analytic treatment. The resonance value, $E_{\mathrm{conv}}=4.3$~MV/cm for $x_{c}=1$~\AA\ (Fig.~1a), is reproduced by the direct integration to better than 1\%.

\subsection*{A2. Harmonic balance and the single-harmonic approximation}
Quartic channel. With $ E=E_{0} \cos\omega t$ , $ F\equiv\left( Z^{*}/m \right) E_{0}$ as in Eq.~\eqref{eq:8}, write $ x=x_{0} \cos\left( \omega t-\phi \right)$ and retain the fundamental. Since ${\cos}^{3}\theta=1/4\left( 3\cos\theta+\cos 3\theta \right)$ with $\theta=\omega t-\phi$ , the cubic restoring force contributes $ 3/4\beta x_{0}^{3}$ to the fundamental. Matching $ \cos\theta$ and $ \sin\theta$ in $\ddot{x}+\gamma\dot{x}+\omega_{\mathrm{TO}}^{2} x+\beta x^{3}=F\cos\omega t$ gives $ x_{0}\left( \Omega-\omega^{2} \right)=F\cos\phi$ and $\gamma\omega\, x_{0}=F\sin\phi$ with $\Omega=\omega_{\mathrm{TO}}^{2}+3/4\beta x_{0}^{2}$ . Squaring and adding gives (8). In phasor form the same step reads $\left[ D\left( \omega \right)+3/4\beta{\left| X \right|}^{2} \right]X=\overset{\sim}{F}$ with $ D\left( \omega \right)=\omega_{\mathrm{TO}}^{2}-\omega^{2}-i\gamma\omega$ and $\left| X \right|=x_{0}$ .

Cubic channel. A quadratic anharmonicity generates no fundamental of its own. It acts by rectifying the motion and by radiating at $ 2\omega$ . Write $ x=c+x_{1} \cos\omega t+x_{2} \cos 2\omega t$ . The dc balance ($\ddot{x}\simeq 0$ , $\omega_{\mathrm{TO}}^{2} c+\alpha\langle x^{2}\rangle=0$) gives $ c=-\alpha x_{1}^{2}/\left( 2\omega_{\mathrm{TO}}^{2} \right)$ . The $ 2\omega$ balance gives $ x_{2}=-\alpha x_{1}^{2}/\left[ 2\left( \omega_{\mathrm{TO}}^{2}-4\omega^{2} \right) \right]$ . The $\omega$ component of $\alpha x^{2}$ is $\alpha\left( 2cx_{1}+x_{1} x_{2} \right)$ . The fundamental therefore obeys $\left[ D\left( \omega \right)+\Delta \right]X=\overset{\sim}{F}$ with

\begin{equation*}
\Delta=-\alpha^{2} x_{0}^{2}\left[ \frac{1}{\omega_{\mathrm{TO}}^{2}}+\frac{1}{2\left( \omega_{\mathrm{TO}}^{2}-4\omega^{2} \right)} \right],  a\left( \omega \right)=\alpha^{2}\left[ \frac{1}{\omega_{\mathrm{TO}}^{2}}+\frac{1}{2\left( \omega_{\mathrm{TO}}^{2}-4\omega^{2} \right)} \right].
\end{equation*}
At $\omega\simeq\omega_{\mathrm{TO}}$ this reduces to $ 5/6\alpha^{2}/\omega_{\mathrm{TO}}^{2}$ , the static offset and the $ 2\omega$ component contributing in the ratio 6:1. Retaining only the rectified displacement returns $\alpha^{2}/\omega_{\mathrm{TO}}^{2}$ and overestimates $ a$ by 20\%. Three qualifications, none of which affects the resonant results. (i) The $ 2\omega$ denominator is written without its damping, $-2i\gamma\omega$ . Restoring it replaces $ x_{2}$ by $ x_{2}/\left( 1+2i/3Q \right)$ at $\omega\simeq\omega_{\mathrm{TO}}$ , shifting the real part of $ a$ by $\sim 0.1\%$ and giving it an imaginary part of $\sim 1.5\%$ at $ Q=9$ , a nonlinear damping term, below the accuracy claimed here. (ii) $ a$ is frequency dependent in this channel. It is $ 3/2\alpha^{2}/\omega_{\mathrm{TO}}^{2}$ as $\omega\to 0$ , changes sign at $\omega=\sqrt{3/8}\,\omega_{\mathrm{TO}}\approx 0.61\,\omega_{\mathrm{TO}}$ , and diverges at $\omega=\omega_{\mathrm{TO}}/2$ , where the second harmonic is resonant. The value $ 5/6$ used in the main text is the near-resonance one, so off-resonance statements about the cubic channel require $ a\left( \omega \right)$ . The quartic entry $ a=3/4\beta$ is frequency independent. (iii) Corrections from the $ 3\omega$ component are relative $ O\left( \xi_{4} \right)$ for the quartic channel (below 1\% at $\xi_{c}$) and relative $ O\left( \xi_{3}^{2} \right)$ for the cubic one. The frequency--amplitude relation $\Delta\omega_{\mathrm{TO}}/\omega_{\mathrm{TO}}=(3/8)\xi_{4}-(5/12)\xi_{3}^{2}$ that follows from these coefficients is the standard one for an oscillator with quadratic and cubic anharmonicity \cite{nayfeh1979}.

\subsection*{A3. Branch points}
Setting $ P\left( u \right)=a^{2} u^{3}\mp 2asu^{2}+\left( s^{2}+G^{2} \right) u-F^{2}$ and $ P'\left( u \right)=3a^{2} u^{2}\mp 4asu+\left( s^{2}+G^{2} \right)=0$ gives (9). The discriminant of $ P'$ is $ 4a^{2}\left( s^{2}-3G^{2} \right)$ , so the roots are real only for $\left| s \right|\geq\sqrt{3} G$ . At $ s=0$ , $ u=\pm iG/\left( \sqrt{3} a \right)$ and $ F^{2}=u\left[ {\left( au \right)}^{2}+G^{2} \right]=\pm i\cdot 2G^{3}/\left( 3\sqrt{3} a \right)$ , giving $\left| F^{2} \right|=2G^{3}/\left( 3\sqrt{3} a \right)$ . Critical values in terms of $\xi$ . At $ s=0$ the radius is $ R=2G^{3}/\left( 3\sqrt{3} a \right)$ , so the linear amplitude at the edge of convergence is $ u_{\mathrm{lin}}=R/G^{2}=2G/\left( 3\sqrt{3} a \right)$ , giving $\xi_{4,c}=\left( 4a/3 \right) u_{\mathrm{lin}}/\omega_{\mathrm{TO}}^{2}=8/\left( 9\sqrt{3} Q \right)$ and $\xi_{3,c}^{2}=\left( 6/5 \right) au_{\mathrm{lin}}/\omega_{\mathrm{TO}}^{2}=4/\left( 5\sqrt{3} Q \right)$ , i.e. Eq.~\eqref{eq:10}. Evaluating $\xi$ instead at the branch-point amplitude $\left| u_{\pm} \right|=G/\left( \sqrt{3} a \right)$ returns values larger by 3/2, namely $ 0.770/Q$ and $ 0.832/\sqrt{Q}$ . Those are the magnitudes at the complex singularity and are not the convergence criterion, which is parameterised by the linear amplitude of Eq.~\eqref{eq:4}. At the saddle-node, $ s=\sqrt{3}\sigma G$ , one has $\left| F^{2} \right|=8\sqrt{3} G^{3}/\left( 9a \right)$ and ${\left| \omega_{\mathrm{TO}}^{2}-\omega^{2}-i\gamma\omega \right|}^{2}=4G^{2}$ , so $ u_{\mathrm{lin}}=2G/\left( 3\sqrt{3} a \right)$ is unchanged. The same $\xi_{c}$ is reached at twice the drive field, as stated in the main text.

Off resonance. Substituting $ u=w/a$ into the cubic gives $ aF^{2}=P\left( w \right)$ with $ P\left( w \right)=w^{3}\mp 2sw^{2}+\left( s^{2}+G^{2} \right) w$ , and the critical points are the roots of $ P'$ , $ w_{\pm}=\left[ \sigma\, 2s\pm\sqrt{s^{2}-3G^{2}} \right]/3$ . Writing $ P\left( w_{\pm} \right)=\left[ \left( 6G^{2}-2s^{2} \right) w_{\pm}+2s\left( s^{2}+G^{2} \right) \right]/9$ and expanding for $\left| s \right|\gg G$ gives $ R_{+}=G^{2}\left| s \right|/a$ at the critical point of larger $\left| u_{\pm} \right|$ , and $ R_{-}=4{\left| s \right|}^{3}/\left( 27a \right)$ at the one of smaller $\left| u_{\pm} \right|$ . Far off resonance the smaller- $\left| u \right|$ critical point always carries the larger $\left| F^{2} \right|$ . Which of the two bounds the series is not decided by these magnitudes but by the sheet on which the physical branch lives. The physical branch is the continuation of $ u=0$ at $ F^{2}=0$ , and it is singular only where it meets another branch. Following that continuation, it reaches the critical point of smaller $\left| u_{\pm} \right|$ first. This is the local maximum of $ P$ , i.e. the saddle-node $ u_{-}$ for $ s>\sqrt{3}\, G$ on the hardening side and $ u_{+}$ for $ s<-\sqrt{3}\, G$ on the softening side. In both cases that point carries $ R_{-}=4{\left| s \right|}^{3}/\left( 27a \right)$ . The remaining critical point, of larger $\left| u_{\pm} \right|$ and $ R_{+}=G^{2}\left| s \right|/a$ , is where the two upper branches are created. For $ 0<F^{2}<R_{+}$ those branches do not yet exist, so the branch issuing from $ u=0$ passes that value analytically and is not bounded by it. (At $ s=+10G$ with $ a=G=1$ this is explicit. $ P$ has its local maximum $ F^{2}=151.5$ at $ u=3.38$ and its local minimum $ F^{2}=10.0$ at $ u=9.95$ , and reverting the series $ u\left( F^{2} \right)$ returns a radius of 153, the maximum, not the minimum.) Hence far off resonance, on both sides, $ E_{\mathrm{conv}}=\left( m/Z^{*} \right)\sqrt{R_{-}}=2\left( m/Z^{*} \right){\left| s \right|}^{3/2}/\left( 3\sqrt{3a} \right)$ , independent of $ G$ and therefore of $ Q$ , which is Eq.~\eqref{eq:12}. The opposite prescription, taking the critical value of smaller modulus irrespective of sheet as an earlier version of this work did, selects the wrong sheet, returns $ Q^{-1}$ on both sides, and is wrong. It would send $ E_{\mathrm{conv}}\to 0$ as $\omega\to 0$ , whereas there it tends to the finite value $\left( 4/9 \right)\left( m\, x_{c}/Z^{*} \right)\omega_{\mathrm{TO}}^{2}$ (about 70 MV/cm for LiNbO$_{3}$). Inside the silent window $\left| s \right|<\sqrt{3}\, G$ the two critical points are complex conjugates of equal modulus, so the ambiguity does not arise and Eq.~(10) holds.

\subsection*{A4. Series reversion}
At resonance the quartic amplitude equation reduces to $ D=y^{3}+y$ with $ y=au/G$ and $ D=aF^{2}/G^{3}$ . Reversion gives $ y=D-D^{3}+3D^{5}-12D^{7}+\cdots$ . Branch points at $ y=\pm i/\sqrt{3}$ give $\left| D \right|=\left| y^{3}+y \right|=2/\left( 3\sqrt{3} \right)=0.385$ .

\subsection*{A5. Multimode algebra}
With $ V_{c}=\lambda X^{2} q$ and damping neglected, $\ddot{X}+\omega^{2} X+\beta_{0} X^{3}+2\lambda Xq=f\cos\omega t$ and $\ddot{q}+\omega_{R}^{2} q+\lambda X^{2}=0$ . With $ X=X_{0} \cos\omega t$ , $ q=-\left( \lambda X_{0}^{2}/2 \right)\left[ 1/\omega_{R}^{2}+\cos 2\omega t/\left( \omega_{R}^{2}-4\omega^{2} \right) \right]$ . The $\omega$ -component of $ 2\lambda Xq$ is $ 2\lambda X_{0}\left( q_{c}+q_{2\omega}/2 \right) \cos\omega t=-\lambda^{2} X_{0}^{3}\left[ 1/\omega_{R}^{2}+1/\left( 2\left( \omega_{R}^{2}-4\omega^{2} \right) \right) \right]\cos\omega t$ . Comparing with the $\omega$ -component of $\beta_{0} X^{3}$ , $\left( 3/4 \right)\beta_{0} X_{0}^{3}$ , gives (13). Damping of $ q$ regulates the $\omega_{R}\to 2\omega$ divergence.

\subsection*{A6. Derivation of Eq.~\eqref{eq:17}}
Quartic. $\xi_{4}\left( E_{\mathrm{ref}} \right)=\left( 8/3 \right) S$ from (15). $\xi_{4}\propto E^{2}$ , so $ E_{\mathrm{conv}}=E_{\mathrm{ref}}{\left[ \xi_{4,c}/\xi_{4}\left( E_{\mathrm{ref}} \right) \right]}^{1/2}=E_{\mathrm{ref}}{\left[ \left( 8/9\sqrt{3} Q \right)\cdot\left( 3/8S \right) \right]}^{1/2}=E_{\mathrm{ref}}{\left[ 1/\left( 3\sqrt{3} QS \right) \right]}^{1/2}$ . 

Cubic. $\left| \xi_{3}\left( E_{\mathrm{ref}} \right) \right|={\left( 12S/5 \right)}^{1/2}$ from (16). $\xi_{3}\propto E$ , so $ E_{\mathrm{conv}}=E_{\mathrm{ref}}\,\xi_{3,c}/\xi_{3}\left( E_{\mathrm{ref}} \right)=E_{\mathrm{ref}}{\left( 4/5\sqrt{3} Q \right)}^{1/2}/{\left( 12S/5 \right)}^{1/2}=E_{\mathrm{ref}}{\left[ 1/\left( 3\sqrt{3} QS \right) \right]}^{1/2}$ . The two routes agree exactly.

\section*{Acknowledgments}
This work was supported by the National Key Research and Development Program of China (2024YFA1409500), the National Natural Science Foundation of China ( 12574373, 12474344), Fundamental Research Funds for the Central Universities (63263125), the 111 Project (B23045).


\section*{Data availability}
No data were generated in this work.

\clearpage
\bibliography{refs}
\end{document}


\title{Supplemental Material for\\
Silent breakdown of the perturbative polarization expansion in the terahertz regime}

\author{Yao Lu}
\email{yaolu@nankai.edu.cn}
\affiliation{Key Laboratory of Weak-Light Nonlinear Photonics, Ministry of Education, School of Physics and TEDA Applied Physics Institute, Nankai University, Tianjin, 300071, China}
\affiliation{Collaborative Innovation Center of Extreme Optics, Shanxi University, Taiyuan, Shanxi, 030006, China}
\author{Jiawei Zhang}
\affiliation{Key Laboratory of Weak-Light Nonlinear Photonics, Ministry of Education, School of Physics and TEDA Applied Physics Institute, Nankai University, Tianjin, 300071, China}

\author{Qiang Wu}
\email{wuqiang@nankai.edu.cn}
\author{Jingjun Xu}
\email{jjxu@nankai.edu.cn}
\affiliation{Key Laboratory of Weak-Light Nonlinear Photonics, Ministry of Education, School of Physics and TEDA Applied Physics Institute, Nankai University, Tianjin, 300071, China}
\affiliation{Collaborative Innovation Center of Extreme Optics, Shanxi University, Taiyuan, Shanxi, 030006, China}

\date{\today}

\maketitle

\numberwithin{equation}{section}
\renewcommand{\theequation}{S\thesection.\arabic{equation}}

\renewcommand{\thefigure}{S\arabic{figure}}
\renewcommand{\thetable}{S\arabic{table}}
\setcounter{figure}{0}
\setcounter{table}{0}

\vspace{4mm}
\stepcounter{section}
\section*{S1. Why the lattice: displacement lever arm, mode-index and geometry dependence, and the reversible window}
\subsection*{S1A. Displacement lever arm: electronic versus phonon-polariton resonance}
The driven displacement that sets the expansion parameter is fixed by the linear response. For a damped driven oscillator of reduced mass m and effective charge $Z^{\ast}$, the steady-state amplitude at drive frequency $\omega$ follows from Eq.~(2):

\begin{equation}
x_{0} = \frac{(Z^{\ast}/m) E}{\left|\omega_{\mathrm{TO}}^{2} - \omega^{2} - i\gamma\omega\right|}.
\label{eq:S1.1}	
\end{equation}
Far below resonance this reduces to $ x_{0} = (Z^{\ast}/m)E/\omega_{\mathrm{TO}}^{2}$, and at exact resonance $(\omega = \omega_{\mathrm{TO}})$ it reduces to $ x_{0} = (Z^{\ast}/m)E/ (\gamma\omega_{\mathrm{TO}}),$ i.e. the off-resonant value multiplied by the quality factor $Q =\omega_{\mathrm{TO}}/\gamma$.

The comparison that underlies the main text is between a lattice (phonon-polariton) mode and a typical electronic resonance, both driven at the same field $E$. Taking the ratio of the two amplitudes of Eq.~\eqref{eq:S1.1} evaluated far below their respective resonances,

\begin{equation}
\frac{x_{0}^{(\mathrm{lat})}}{x_{0}^{(\mathrm{el})} }= \frac{Z^{\ast}_{\mathrm{lat}}/{m_{\mathrm{lat}}}}{Z^{\ast}_{\mathrm{el}} / m_{\mathrm{el}}} \times \frac{\omega_{\mathrm{el}}^2}{\omega_{\mathrm{TO}}^{2}}.
\label{eq:S1.2}
\end{equation}
With the lattice parameters of Table~\ref{tab:S1} $(Z^{\ast}_{\mathrm{lat}} \approx 5e, m_{\mathrm{lat}} \approx 34.6 \text{ u})$\cite{lu2021} and a representative electronic resonance $(Z^{\ast}_{\mathrm{el}} = e, m_{\mathrm{el}} = m_{e}, \omega_{\mathrm{el}} \approx 1.1\times 10^{16} $ s$^{-1}),$ Eq.~\eqref{eq:S1.2} gives

\begin{equation}
\frac{x_{0}^{(\mathrm{lat})}}{x_{0}^{(\mathrm{el})} } \approx 4.5,
\label{eq:S1.3}
\end{equation}
which is the factor quoted in the main text at fixed field. Driving the lattice mode at its own resonance adds a further factor $\omega_{\mathrm{TO}}/\gamma = Q \approx$ 9, so that

\begin{equation}
\frac{x_{0}^{(\mathrm{lat})}}{x_{0}^{(\mathrm{el})} }\approx 4.5 \times Q \approx 40
\label{eq:S1.4}
\end{equation}
Because $\xi$ is quadratic in the driven displacement, $\xi = a x_{0}^{2}/\omega_{\mathrm{TO}}^{2}$ (Eq.~(5)), the same lever arm enters $\xi$ squared. It is this quadratic amplification, and not the intrinsic strength of the lattice anharmonicity, that allows a phonon-polariton resonance to reach its convergence boundary at a field far below the electronic damage scale. The numerical values in Eqs.~\eqref{eq:S1.3} and \eqref{eq:S1.4} depend on the reference electronic parameters; the set given above reproduces the factors quoted in the main text.

Two qualifications apply. First, the comparison presumes comparable oscillator strengths; the residual difference is a material ratio that we do not evaluate, so Eqs.~\eqref{eq:S1.3} and \eqref{eq:S1.4} are lever arms for the driven displacement and not statements about the relative size of the susceptibilities. Second, the lattice value is quoted for the bare transverse-optical frequency, i.e. the deeply confined ($n \to \infty)$ limit of Sec. S1B, which carries the stiffest restoring force. On a propagating polariton branch the displacement per unit internal field is larger, so the leverage is if anything underestimated and the convergence field correspondingly lower.

\subsection*{S1B. Mode index, geometry, and the delocalized limit; relation to Ref. \cite{lu2021}}
In a structured medium the nonlinearity extracted from a measurement is not the bulk coefficient but an effective one, obtained by projecting the driven polarization onto the detected mode. For modes indexed by $i$ and $j$,

\begin{equation}
\chi_{\mathrm{eff}}^{(2)} = \Gamma_{ij} \chi^{(2)}, \Gamma_{ij} = \int \frac{\chi^{(2)}(z) E_{i}^{*}(z)  E_{j}(z)}{N_{i} N_{j}} dz,
\label{eq:S1.5}
\end{equation}
with $ N_{i} = (\int |E_{i}|^{2} dz / n^{2})^{1/2}$ the mode normalization; $\chi_{\mathrm{eff}}^{(3)}$ carries one further overlap factor. Two consequences follow for the present argument.

{\bf{(i) Geometry dependence.}} $\chi_{\mathrm{eff}}$ for a thin-film waveguide differs from the bulk value by the overlap factor, typically 0.3$\sim$0.8 for a 1$\sim$2 $\upmu$m film at terahertz frequencies. Absolute coefficients quoted for different samples are therefore not directly comparable without a common normalization.

{\bf{(ii) Cancellation in the criterion.}} The convergence criterion, Eq.~(17) of the main text, is expressed through the fractional mode shift $S = \left|\delta\omega_{\mathrm{TO}}/\omega_{\mathrm{TO}}\right|,$ an intrinsic quantity independent of waveguide geometry and of the probe path. The internal field is related to the incident field by $ E_{\mathrm{int}} = F(\omega) E_{\mathrm{inc}}$, with $ F(\omega)$ of order unity in bulk and 2$\sim$10 in a microcavity; since Eq.~(17) is evaluated at the internal reference field, $ F(\omega)$ enters only as a conversion factor and does not move the convergence radius. This is why the criterion can be applied across samples and geometries without recalibration.

\vspace{3 mm}
{\bf{Renormalized restoring force, and the relation to Ref. \cite{lu2021}}}

Reference \cite{lu2021} writes the nonlinear Huang equations with two fields: an external drive $ E_{T}$ acting directly on the ions, and the macroscopic field $ E$ that the ions radiate, which is eliminated by assuming that the polariton field propagates with $ k^{2} = \varepsilon_{0}\omega^{2}/c^{2}$ . Our Eqs.~(2)-(3) instead take $ E$ to be the total macroscopic field, so no elimination is needed: Maxwell's equations then fix only the dispersion, $ k^{2}(\omega) = \varepsilon\left( \omega \right)\omega^{2}/c^{2}$ , and leave the amplitude free, which is why Eq.~(4) follows directly. The two bookkeepings differ in what the symbol $ E$ denotes; it is nevertheless useful to carry out the elimination for a general mode index $ n^{2} \equiv k^{2} c^{2}/\omega^{2}$ . From Eq.~(3) and the transverse wave equation, $(n^{2} - \varepsilon_{\infty}) E=NZx/\varepsilon_{0}$ , and with $ NZ^{2}/(\epsilon_{0} m) = \omega_{\mathrm{TO}}^{2}\Delta\varepsilon$ , $\Delta\varepsilon \equiv \varepsilon_{0}-\varepsilon_{\infty}$ , the equation of motion acquires a renormalized restoring force $\omega_{\mathrm{eff}}^{2}=\omega_{\mathrm{TO}}^{2}\left[ 1-\Delta\varepsilon/\left( n^{2}-\varepsilon_{\infty} \right) \right]$ and $ D_{\mathrm{eff}}=\omega_{\mathrm{eff}}^{2}-\omega^{2}-i\gamma\omega=D_{0}\left( n^{2}-\varepsilon\left( \omega \right) \right)/\left( n^{2}-\varepsilon_{\infty} \right)$ , with $ D_{0}=\omega_{\mathrm{TO}}^{2}-\omega^{2}-i\gamma\omega$ . Four limits fix the interpretation: $ n^{2} \to \infty$ (deeply confined mode) gives $\omega_{\mathrm{eff}}\to\omega_{\mathrm{TO}}$ , the bare ionic response of Eq.~(4); $ n^{2}=\varepsilon_{0}$ gives $\omega_{\mathrm{eff}}= 0$ , the Drude form of Ref. \cite{lu2021}; $ n^{2} \to 0$ gives $\omega_{\mathrm{eff}}^{2}= \omega_{\mathrm{TO}}^{2}\varepsilon_{0}/\varepsilon_{\infty}= \omega_{\mathrm{LO}}^{2}$ by Lyddane-Sachs-Teller; and $ n^{2} = \varepsilon\left( \omega \right)$ gives $ D_{\mathrm{eff}}= 0$ , the polariton shell, i.e. the phase-matching pole. The interval $\varepsilon_{\infty}< n^{2}< \varepsilon_{0}$ , where $\omega_{\mathrm{eff}}^{2}< 0$ , is the Reststrahlen band.

$E$ is the total Born-Huang field. The external drive enters only through Maxwell's equations and the boundary conditions, not as a separate force. Writing a second term $ qE_{\mathrm{ex}}/m$ alongside the macroscopic-field term, as Ref. \cite{lu2021} does, is legitimate only if one of the two fields is then eliminated with the wavevector as an input, the step analyzed above, which returns Eq.~(4) as $ n^{2}\to\infty$ . Eq.~(3) fixes the observable: $ P\left( E \right)=NZ^{*} x\left( E \right)+\epsilon_{0}\left( \varepsilon_{\infty}-1 \right) E$ , and its second term is entire in $ E$ , so $ P$ and $ x$ share the same singularity structure and the same convergence radius; the radius is a property of the local constitutive relation Eq. (2) and is expressed in the internal field. Closing Eqs. (2)-(3) with Maxwell's equations makes the problem polariton: for a transverse mode of index $ n^{2}=k^{2} c^{2}/\omega^{2}$ , $\left( n^{2}-\varepsilon_{\infty} \right) E=NZ^{*} x/\epsilon_{0}$ , and eliminating $ E$ from Eq. (2) leaves a free nonlinear oscillator with the renormalized restoring force $\omega_{\mathrm{eff}}^{2}\left( n^{2} \right)$ above, the drive entering only as a boundary condition. Consequently the main text, evaluated at $ n^{2}\to\infty$ where $\omega_{\mathrm{eff}}=\omega_{\mathrm{TO}}$ is the stiffest restoring force, quotes a conservative bound on the internal field. On a propagating branch ($\omega_{\mathrm{eff}}<\omega_{\mathrm{TO}}$) the displacement per unit internal field is larger, and at resonance $ E_{\mathrm{conv}}\propto{\left( \gamma\omega_{\mathrm{eff}} \right)}^{3/2}$ is lower by ${\left( \omega_{\mathrm{eff}}/\omega_{\mathrm{TO}} \right)}^{3/2}$ , a factor set by waveguide thickness, mode order and geometry, not only frequency, the polariton-specific prediction absent from Ref. \cite{lu2021}.

Two observations bear on how the literature is read. (i) The vanishing of the restoring force is the single point $ n^{2}= \varepsilon_{0}$ , the $\omega\to 0$ limit of the polariton dispersion; it is not a generic polariton property. (ii) The Drude denominator $-\omega^{2}-i\gamma\omega$ then carries no restoring force, so $\chi^{\left( 1 \right)}\propto{\left( \omega^{2}+i\gamma\omega \right)}^{-1}$ diverges as $\omega\to 0$ , and the higher orders built from it do too: the approximation and its result fail at the same end. Because $ D_{\mathrm{eff}}$ is a near-cancellation of two large terms it is also hypersensitive to $ n^{2}$ , with the parameters of Ref. \cite{lu2021}, a $\sim 1\%$ change in $ n^{2}$ shifts $\left| D_{\mathrm{eff}} \right|$ by order unity and the nonlinearity built from it by an order of magnitude. This predicts a strong, directly testable dependence of the effective nonlinearity on the driven mode index (waveguide thickness, mode order, geometry), not contained in Ref. \cite{lu2021}. None of it alters the local algebra of the main text. It only fixes the conversion to the incident field and, through $\omega_{\mathrm{eff}}\left( n^{2} \right)$ , how conservative the quoted bound is.

\subsection*{S1C. The reversible strongly nonlinear window}
Above the convergence field and below the threshold for irreversible change, the lattice is strongly and nonlinearly driven yet returns to its initial state once the field is removed. Inside this window the polarization expansion has failed, every truncated $\chi(n)$ description is quantitatively wrong, while the crystal survives intact; the failure is a statement about the expansion, not about the material. The window is bounded by two curves.

Lower boundary, $ E_{\mathrm{conv}}(\omega),$ from Eq.~(9). With the parameters of S3 it has a minimum of 4.3 MV/cm at 7.6 THz (taking $ x_{c} = 1 \text{\AA})$ and rises steeply on either side, as $\left|\omega^{2} - \omega_{\mathrm{TO}}^{2}\right|^{3/2}$ outside the silent sub-band. At the edges of the silent sub-band (6.9 and 8.3 THz) it reaches $\approx$ 8.6 MV/cm, and at 5.8 and 8.8 THz it reaches $\approx$ 40 MV/cm.

Upper boundary, the threshold for irreversible change. For LiNbO$_3$ over the terahertz range this is 10$\sim$100 MV/cm, the lower edge set by the observation of fifth-harmonic generation without damage \cite{hoegen2018} and the upper edge by the nonthermal-melting threshold \cite{henstridge2022}.

The reversible nonperturbative window is the region $ E_{\mathrm{conv}} <$ E $< E_{\mathrm{damage}}$, i.e. 5.8$\sim$8.8 THz here. Within it, the silent sub-band $\left|s\right| <\sqrt{3}G$ (6.9$\sim$8.3 THz) is the range in which no jump or multistability can occur. The window is entered before the structural instability and nonthermal melting of Refs. \cite{henstridge2022,feng2025,li2019}, which set the upper boundary.

\vspace{3 mm}
{\bf{Why the lattice is the natural host}}

In the terahertz range the drive necessarily lies inside or immediately adjacent to the Reststrahlen band, so the resonance cannot be avoided: absorption, the accompanying Kramers-Kr{\"o}nig fluctuation-dissipation noise floor, and heating under continuous-wave drive are bundled with the giant nonlinearity, and a lossless, noiseless terahertz nonlinearity is thermodynamically excluded. This is not a defect of the platform but the reason the question can be posed at all: because leaving the perturbative range here means structural instability rather than ionization, the reversible nonperturbative window opens before any damage.

\stepcounter{section}
\section*{S2. The expansion parameter: asymptotic form and two misreadings}
\subsection*{S2A. Transfer theorem and the 1/n correction}
For a function whose nearest singularity in the complex plane is a branch point of exponent $\kappa$, the transfer theorem for algebraic singularities fixes the large-n behaviour of the Taylor coefficients. Applied to the polarization series, whose coefficients are generated by the repeated insertion of one anharmonic vertex per order, it gives

\begin{equation}
c_{n} \sim \xi^{n} n^{-\kappa-1} / \Gamma(-\kappa),
\label{eq:S2.1}
\end{equation}
so that the ratio of successive terms is

\begin{equation}
\left|c_{n} / c_{n-1}\right| = \xi [1 - (\kappa+1)/n + O(1/n^{2})].
\label{eq:S2.2}
\end{equation}
For the fold (saddle-node) singularity the amplitude relation (8) is a square-root branch point, $\kappa = 1/2$, so the correction is $3/(2n)$:

\begin{equation}
\left|c_{n} / c_{n-1}\right| = \xi [1 - 3/(2n) + O(1/n^{2})].
\label{eq:S2.3}
\end{equation}

Dividing by the critical value $\xi_c$ of Eq.~(10) of the main text, the normalized ratio $|c_n/c_{n-1}|/\xi_c$ tends to unity as $n\to\infty$, which is the form quoted
in Eq.~(6) of the main text; Eqs.~\eqref{eq:S2.2} and \eqref{eq:S2.3} use the unnormalized ratio. The correction vanishes as $n\to \infty$ and is below 30\% for $n\geq 5$, so $\xi = a x_{0}^{2}/\omega_{\mathrm{TO}}^{2}$ is the effective expansion parameter to within the model precision. Its critical value is not set by this asymptotic ratio but by the exact location of the singularity, Eq.~(9). At resonance $\xi_{c} = 8/(9\sqrt 3 Q) = 0.513/Q$ in the $\beta$-dominated (quartic) channel and $0.680/\sqrt{Q}$ in the $\alpha$-dominated (cubic) channel. At $Q = 9$ the series already fails near $\xi \approx 6\times 10^{-2}$, an order of magnitude below the naive value $\xi = 1$.

The behaviour is explicit for the resonant quartic channel of Fig. 2(a). Reversion of the amplitude equation by Lagrange inversion gives successive term ratios 1, 3, 4, 4.58, 4.96, 5.23, $\ldots$, converging to 27/4 with an $O(1/n)$ correction, the limit being the radius in the variable of Fig. 2(a). The ratio therefore rises monotonically towards its limit, so low-order truncations systematically look better converged than the asymptotic rate predicts: the failure is not announced at the order at which it sets in.

\subsection*{S2B. Two misreadings that flatter the series}
Evaluating $\xi$ at the branch-point amplitude $\left| u_{\pm} \right|$ rather than at the linear amplitude of Eq.~(4). The two differ by a factor of 3/2 at resonance, and using the branch-point amplitude overestimates the convergence field. The expansion parameter orders successive terms of the series generated by iterating Eq.~(2) about the linear solution, so it is the linear amplitude that sets the ordering; the branch-point amplitude is where the series finally fails, not the quantity that orders its terms.

Taking the convergence radius to be the critical value of smallest modulus. As shown in Methods A3, the radius is set by the critical point reached by continuation along the physical branch, which far off resonance is the one of larger $\left| F^{2} \right|$ . Both misreadings go in the same direction: they make the truncated series look better converged than it is, which is precisely why the breakdown is silent.

\stepcounter{section}
\section*{S3. Material parameters, the anchoring interval $ x_{c}$, and the three caveats}

\vspace{3 mm}
{\bf{Parameters}}

For the LiNbO$_3$ A$_1$ polariton mode the parameters used throughout are collected in Table~\ref{tab:S1}.

\begin{table}[tb]
\centering
\caption{Parameters for the LiNbO$_3$ A1 polariton mode.\label{tab:S1}}
\begin{tabular}{lcc}
\hline
Parameter & Symbol & Value \\
\hline
A1 TO frequency & $\omega_{\mathrm{TO}}/2\pi$ & 7.6 THz \\
Damping & $\gamma/2\pi$ & 0.84 THz \\
Quality factor & Q = $\omega_{\mathrm{TO}}/\gamma$ & 9.0 \\
Reststrahlen ratio & $\omega_{\mathrm{LO}}/\omega_{\mathrm{TO}}$ & 1.46 \\
Static / high-frequency ratio & $\varepsilon_0/\varepsilon_\infty$ & 2.1 \\
Reduced mass & $m_{r}$ & $\approx$ 34.6 u \\
Effective charge & $Z^{\ast}$ & $\approx$ 5e \\
Anchoring interval & $x_{c}$ & 0.5--2 $\text{\AA}$ \\
Fractional shift at 1 MV/cm & $S$ & $2.8\times 10^{-4}\sim4.5\times 10^{-3}$ \\
\hline
\end{tabular}
\end{table}

\vspace{3 mm}
{\bf{Anchoring interval for $ x_{c}.$}}

The displacement $ x_{c} = \omega_{\mathrm{TO}}/\sqrt\beta$, at which the quartic restoring force equals the harmonic one, is not measured directly. It is anchored to the driven-displacement scale reported in Ref. \cite{hoegen2018}, where a different LiNbO$_3$ mode $(\approx 19\,\mathrm{THz})$ is driven into the anharmonic regime; the interval $0.5\sim2 \text{\AA}$ is deliberately wide, so as to absorb the difference between modes, the uncertainty in the mode Gr{\"u}neisen parameter, and the contribution of sixth-order terms to the branch point. The dimensionless result of Eq.~(9) is converted to an absolute field through $ x_{c}$, which is the origin of the entry of Table I. The resulting spread in Table I, $E_{\mathrm{conv}}=2.2\sim9.0$ MV/cm, is therefore a systematic uncertainty band, not a confidence interval: it reflects the deliberately conservative width of the anchoring interval and would narrow proportionally if $x_c$ were pinned by an independent measurement of the mode Gr\"uneisen parameter or of the driven displacement.

The convergence field at resonance then follows from Eq.~(17),

\begin{equation}
E_{\mathrm{conv}} = E_{\mathrm{ref}} (1 / (3\sqrt 3 Q S))^{1/2},
\label{eq:S3.1}
\end{equation}
with $ E_{\mathrm{ref}} = 1$ MV/cm. For $Q = 9$ and $S = 2.8\times 10^{-4}$ to $ 4.5\times 10^{-3}$ this gives $ E_{\mathrm{conv}} = 2.2\sim9.0$ MV/cm, the range quoted in the main text.

\vspace{3 mm}
{\bf{The three caveats}}

They act in different directions and must be read together.

{\bf{(i) Geometry.}} Wherever the microcavity or waveguide geometry enhances the internal field over the incident field, the tabulated values are upper bounds on the incident field required to reach the boundary.

{\bf{(ii) Damping.}} A decay rate growing with phonon occupancy raises $\gamma$, lowers $Q$, and raises $ E_{\mathrm{conv}}$, so the tabulated values are conservative lower bounds. A Klemens-type growth $\gamma = \gamma_{0}(1 + \eta x_{0}^{2}/a^{2})$ with $\eta \approx$ 1$\sim$10 elevates $\gamma$ by at most $\eta \xi_{c} \approx$ 0.6 at $\xi = \xi_{c} = 0.057$, raising $ E_{\mathrm{conv}}$ by at most $(1.6)^{3/2} \approx$ 2.

{\bf{(iii) Multimode coupling.}} For the LiNbO$_3$ spectrum, no $A_1$-symmetry partner mode lies inside the suppression window (the nearest higher infrared-active mode, near 13 THz, is of $E$-symmetry and decouples from the driven $A_1$ mode through Eq.~(13)), and thus the single-mode values are therefore again conservative lower bounds for LiNbO$_3$. For a general material, where a mode may fall inside the suppression window, they become upper bounds. The multimode correction is controlled by $\kappa = \lambda^{2}/(\beta_{0} \omega_{R}^{2});$ the single-mode estimate is reliable for $\kappa \ll$ 1 and becomes unreliable once $\kappa \gtrsim$ 1, at which point $\beta_{\mathrm{eff}}$ must be evaluated from Eq.~(13). 

\vspace{3 mm}
{\bf{Estimate of the multimode parameter $\kappa$}}

The quantitative reliability of the single-mode bound against multimode coupling rests on $\kappa = \lambda^{2}/(\beta_{0} \omega_{R}^{2}),$ with $\lambda$ the cross-coupling of Eq.~(13) and $\omega_{R}$ the partner-mode frequency. We estimate $\lambda$ from the scale of the reconstructed $ \mathrm{LiNbO}_{3}$ potential of Ref. \cite{hoegen2018}, writing it as a fraction f of the on-site anharmonicity carried by the off-diagonal force constant (f $\sim$ 0.1 for inter-mode couplings). For the highest $ \mathrm{LiNbO}_{3}$ infrared-active mode $(\omega_{R}/2\pi \approx 8\sim9 \mathrm{ THz})$ this gives $\kappa \sim 10^{-2}$; the bracket in Eq.~(13) is then of order unity, so the correction to $\beta_{\mathrm{eff}}$ is at the few-percent level and the single-mode values of Table I are quantitatively reliable to that precision, not merely bounds. Crucially, since no partner mode lies in the suppression window, every coupling only raises $ E_{\mathrm{conv}}$ whatever the value of $\kappa$, so the conservative character of the single-mode numbers does not depend on this estimate. A precise $\kappa$ awaits the full force-constant fit of Ref. \cite{hoegen2018}.

\stepcounter{section}
\section*{S4. Competing bounds: ionization, and the $Q$-scaling in monolayer hBN}
\subsection*{S4A. Keldysh parameter and ionization across the window}
The Keldysh parameter is

\begin{equation}
\gamma_{K} = \omega (2 m^{\ast} \Delta)^{1/2} / (e E).
\label{eq:S4.1}
\end{equation}
With a gap $\Delta = 4$ eV and an effective mass $m^{\ast} = 0.2 m_{e}, \gamma_{K} = 1$ falls at

\begin{equation}
E = \omega(2m^{\ast}\Delta)^{1/2}/e \approx 1.4\,\mathrm{MV}/\mathrm{cm}.
\label{eq:S4.2}
\end{equation}
Across the reversible window ($E \approx 2\sim9$ MV/cm) the Keldysh parameter runs from about 0.7 to 0.2, so the system is formally inside the tunnelling regime.

Ionization is not, however, governed by $\gamma_{K}$ alone; it is controlled by the tunnelling action W, the exponent of the Landau-Dykhne ionization rate $\exp{(-W)}$:

\begin{equation}
W = (4\sqrt 2/3) {m^{\ast}}^{1/2} \Delta^{3/2} / (e \hbar E) = A / E, A = 2.44\times 10^{10} \text{ V/m}.
\label{eq:S4.3}
\end{equation}
Across the window W falls only from $\approx 1.1\times 10^{2}$ to $\approx$ 27, so that $\exp{(-W)}$ remains between about $ 10^{-48}$ and $ 10^{-12}$: ionization is exponentially frozen. W reaches unity only near E $\approx 2\times 10^{2}$ MV/cm, above the damage threshold. Free carriers are therefore not a necessary feature of the nonperturbative regime, and the window describes bound lattice motion. This is why the Keldysh line does not compete with the convergence boundary of Eq.~(9).

\subsection*{S4B. Consistency check in monolayer hBN}
The $ Q^{-3/2}$ scaling of Eq.~(12) can be cross-checked against a second system with a very different quality factor. In monolayer hexagonal boron nitride the in-plane optical phonon has a far longer lifetime than in LiNbO$_3$, so that $ Q_{\mathrm{hBN}}/Q_{\mathrm{LN}} \approx$ 10$\sim$100. Applying

\begin{equation}
E_{\mathrm{conv}}(\mathrm{hBN}) = E_{\mathrm{conv}}(\mathrm{LN}) \times (Q_{\mathrm{hBN}} / Q_{\mathrm{LN}})^{-3/2}
\label{eq:S4.4}
\end{equation}
to the $ \mathrm{LiNbO}_{3}$ range 2$\sim$9 MV/cm gives $ E_{\mathrm{conv}}(\mathrm{hBN}) \approx$ 0.2$\sim$28 MV/m, where the lower corner corresponds to the largest quality-factor ratio $(\approx 100)$ combined with the smallest $ \mathrm{LiNbO}_{3}$ convergence field, and the upper corner to the smallest ratio $(\approx 10)$ combined with the largest $ \mathrm{LiNbO}_{3}$ convergence field.

First-principles calculations for monolayer hBN predict a threshold field of $\approx$ 24 MV/m for order-unity Kerr response and harmonic generation \cite{iyikanat2021}. The predicted convergence field brackets this value rather than lying cleanly below it: it falls below 24 MV/m over most of the range and exceeds it only in the corner set jointly by the smallest $Q$ ratio and the largest $ \mathrm{LiNbO}_{3}$ field. This is consistent with the intended ordering, the convergence boundary marks where the expansion fails, whereas the first-principles threshold marks where the nonperturbative response has grown to order unity, but, because both entries carry an order-of-magnitude uncertainty, from the anchoring interval of S3 and from the $Q$ ratio respectively, the comparison should be read as establishing that the $ Q^{-3/2}$ law extends to a system with a much larger $Q$, not as a quantitative prediction of the hBN threshold. By contrast, in the electronic epsilon-near-zero regime probed in Refs. \cite{reshef2017,reshef2019} the driven displacement is small on the lattice scale, so the convergence field there lies far above the damage threshold; that regime and the present one are distinguished by the displacement lever arm of Sec. S1A.

\stepcounter{section}
\section*{S5. Detecting the breakdown: readout amplification and phase/amplitude discrimination}
\subsection*{S5A. Readout amplification near the Reststrahlen edge}
\begin{figure}[!t]
\centering
\includegraphics[width=0.95\columnwidth]{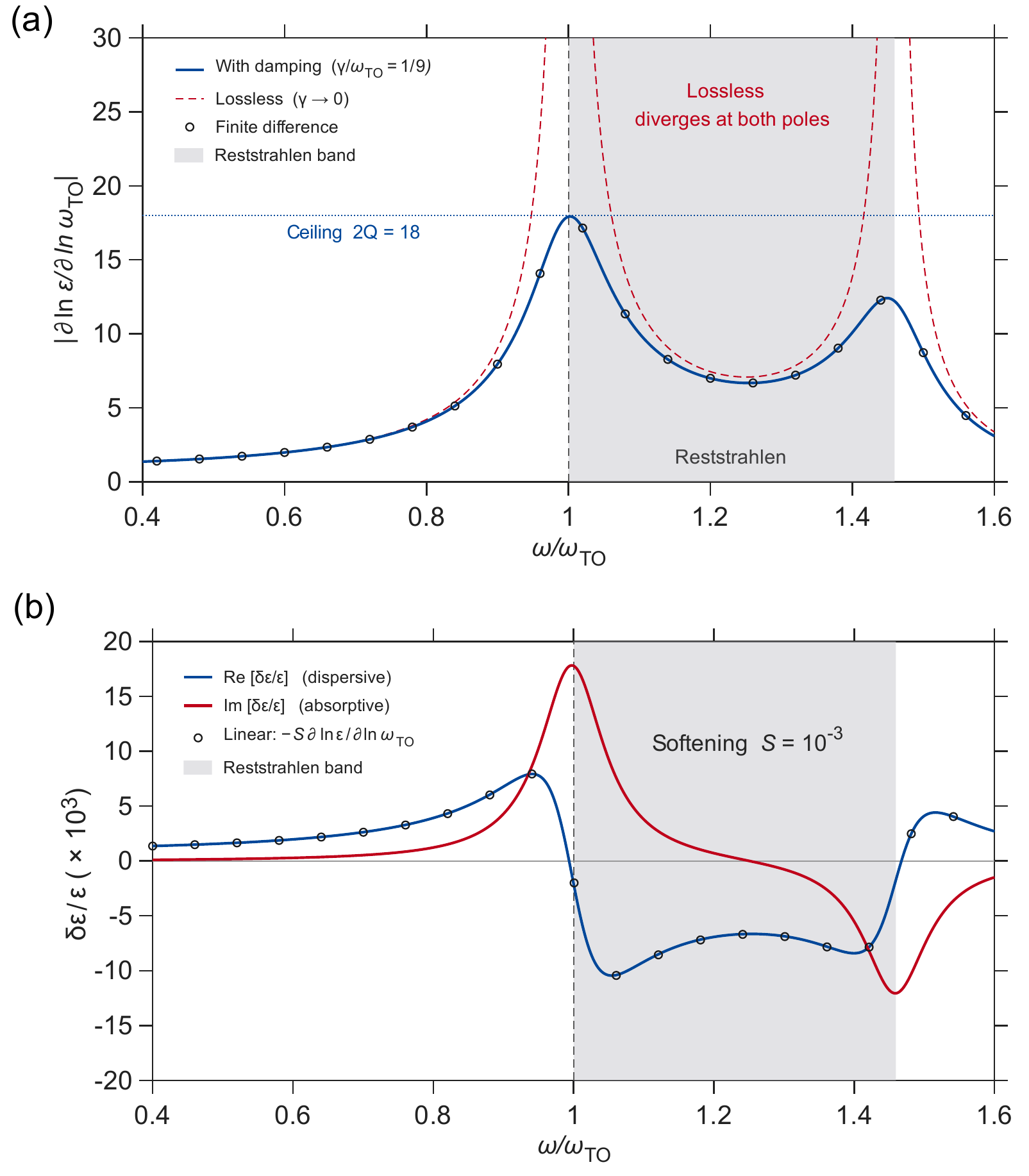}
\caption{Readout amplification near the Reststrahlen edge, with the $ \mathrm{LiNbO}_{3}$ parameters of Table~\ref{tab:S1} $(\omega_{\mathrm{LO}}/\omega_{\mathrm{TO}} = 1.46,\ Q = 9).$
(a) Magnitude of $ \partial \ln \varepsilon/\partial \ln \omega_{\mathrm{TO}}$, Eq.~\eqref{eq:S5.2}, solid line. Open circles evaluate the same derivative by finite difference under the identical constraint, $ \omega_{\mathrm{LO}}^{2}-\omega_{\mathrm{TO}}^{2}$ fixed and $ \gamma$ constant. Dashed curve, the lossless form. Horizontal line, the ceiling $ 2Q = 18$. Shaded, the Reststrahlen band.
(b) Fractional change of $ \varepsilon$ for a softening $ S=10^{-3}$, obtained by direct substitution of the shifted frequencies with no differentiation. Real part, dispersive readout. Imaginary part, absorptive. Circles give the linear prediction $ -S\,\partial\ln\varepsilon/\partial\ln\omega_{\mathrm{TO}}$.\label{fig:S1}}
\end{figure}

Near the Reststrahlen edge the dielectric function is strongly dispersive, so a small fractional shift of the mode frequency produces a large change in $\varepsilon$. Both sides of this readout are fractions. The softening enters as $S=\delta\omega_{\mathrm{TO}}/\omega_{\mathrm{TO}}$, and what a probe at fixed frequency measures is $\delta\varepsilon/\varepsilon$, equivalently $\delta n/n=\frac{1}{2}\,\mathrm{Re}[\delta\varepsilon/\varepsilon]$. The amplification factor is the ratio of the two, that is, the logarithmic derivative, and only in this dimensionless form does the sensitivity saturate at the material-independent ceiling $2Q$ quoted in the main text. With the single-oscillator form

\begin{equation}
\varepsilon(\omega) = \varepsilon_\infty \frac{\omega_{\mathrm{LO}}^{2} - \omega^{2} - i\gamma\omega}{\omega_{\mathrm{TO}}^{2} - \omega^{2} - i\gamma\omega},
\label{eq:S5.1}
\end{equation}
The anharmonic shift is a change of the restoring force at fixed Born charge, so within Eqs.~(1) to (3) the invariant is the oscillator strength $\Omega_{p}^{2}=NZ^{*2}/(\epsilon_{0}m)$, not $\omega_{\mathrm{LO}}$. The LST relation therefore has to be used in the form $\omega_{\mathrm{LO}}^{2}=\omega_{\mathrm{TO}}^{2}+\Omega_{p}^{2}/\varepsilon_{\infty}$, in which the two poles shift by the same additive amount, $d\omega_{\mathrm{LO}}^{2}/d\omega_{\mathrm{TO}}^{2}=1$. Holding $\varepsilon_{0}$, and with it the ratio $\omega_{\mathrm{LO}}^{2}/\omega_{\mathrm{TO}}^{2}$, fixed is a different and equally incorrect choice. Differentiating Eq.~\eqref{eq:S5.1} with respect to $\omega_{\mathrm{TO}}$ under this constraint and dividing by $\varepsilon$ leaves

\begin{equation}
\frac{\partial \ln \varepsilon}{\partial \ln \omega_{\mathrm{TO}}}
=-\frac{2\omega_{\mathrm{TO}}^{2}\left( \omega_{\mathrm{LO}}^{2}-\omega_{\mathrm{TO}}^{2} \right)}
{\left( \omega_{\mathrm{LO}}^{2}-\omega^{2}-i\gamma\omega \right)\left( \omega_{\mathrm{TO}}^{2}-\omega^{2}-i\gamma\omega \right)},
\label{eq:S5.2}
\end{equation}
with $Q=\omega_{\mathrm{TO}}/\gamma$. At the transverse frequency, $\omega=\omega_{\mathrm{TO}}$, the magnitude is

\begin{equation}
\left|\frac{\partial \ln \varepsilon}{\partial \ln \omega_{\mathrm{TO}}}\right|_{\omega=\omega_{\mathrm{TO}}}=\frac{2Q}{\sqrt{1+\left[ Q\left( \omega_{\mathrm{LO}}^{2}/\omega_{\mathrm{TO}}^{2}-1 \right) \right]^{-2}}},
\label{eq:S5.4}
\end{equation}
which for $\mathrm{LiNbO}_{3}$ ($\omega_{\mathrm{LO}}/\omega_{\mathrm{TO}}=1.46$, Q $\approx$ 9) is 17.9, that is, $2Q$ to within 0.5\%. Neglecting the motion of the longitudinal pole altogether, equivalently taking the longitudinal pole far from the transverse one, gives back the compact limiting case quoted alongside Eq.~(18) of the main text, $-2\omega_{\mathrm{TO}}^{2}/(\omega_{\mathrm{TO}}^{2}-\omega^{2}-i\gamma\omega)$.
Near the edge the sensitivity falls off over the linewidth, and in the compact limit, writing $\omega = \omega_{\mathrm{TO}} + \delta$, the magnitude becomes $ 2\omega_{\mathrm{TO}}/\sqrt{4\delta^{2}+\gamma^{2}(1+\delta/\omega_{\mathrm{TO}})^{2}}$. In the full expression~\eqref{eq:S5.2} the amplifier stays within a factor of 1.5 of the ceiling for detunings up to half a linewidth and within a factor of two and a half up to a full linewidth, so the usable optimum lies within of order $\gamma$ of the transverse frequency (Fig.~\ref{fig:S1}). For LiNbO$_3$ (Q $\approx$ 9) the ceiling is 17.9, enough to render a displacement of order $ 10^{-2} \text{\AA}$ measurable. The algebra can be bypassed altogether. In Fig.~\ref{fig:S1}(a) the closed form~\eqref{eq:S5.2} is overlaid with a direct finite-difference evaluation of the same derivative under the identical constraint, and the two are indistinguishable. Figure~\ref{fig:S1}(b) gives the readout itself, and this exact result tracks the linear prediction of Eq.~\eqref{eq:S5.2} to better than $ 2\%$ over the whole band, reaching $ |\delta\varepsilon/\varepsilon|\approx 1.8\times10^{-2}$ at the transverse frequency, that is, $ \delta n/n\approx 10^{-3}$.

A second amplifier is available at the LO frequency, $\omega \to \omega_{\mathrm{LO}}^{-}$, where $\varepsilon \to 0$ in the undamped model. Finite damping and the remaining infrared-active modes cut this divergence off, giving an enhancement of order 10$\sim$100, and it should be treated as a qualitative amplifier rather than a calibrated factor. Far below the edge, as in Ref. \cite{huang2024}, 0.6 THz against $\omega_{\mathrm{TO}}/2\pi = 7.6$ THz, the magnitude of Eq.~\eqref{eq:S5.2} is only $\approx$ 1.1, whereas the fixed-$\omega_{\mathrm{LO}}$ form would give 2, so the readout there is suppressed rather than absent. Evaluated at $\omega=\omega_{\mathrm{LO}}$ the same expression gives $\approx 12$ for these parameters, within the range quoted above. The same caveat attaches to the lossless curve of Fig.~\ref{fig:S1}(a), which diverges at both poles and therefore overstates the amplifier by close to an order of magnitude at this $ Q$. Its $ \varepsilon \to 0$ limit is a natural epsilon-near-zero point, sharp only because damping and the remaining infrared-active modes are omitted.

\FloatBarrier

\subsection*{S5B. Phase/amplitude discrimination in Measurements (2)}
At exact resonance the linear response is in quadrature with the drive, $ x_{0}(1) = i F/(\gamma\omega_{\mathrm{TO}}),$ while the anharmonicity adds a real frequency shift $\delta\omega_{\mathrm{TO}}$, i.e. a component in phase with the drive. Writing the response with the shifted frequency, $ x_{0} = F/(2\omega_{\mathrm{TO}} \delta\omega_{\mathrm{TO}} - i\gamma\omega_{\mathrm{TO}}),$ and using $\delta\omega_{\mathrm{TO}}/\omega_{\mathrm{TO}} = S$,

\begin{equation}
x_{0} = \frac{x_{0}(1)}{1 + 2iQS}.
\label{eq:S5.5}
\end{equation}
Expanding for small $S$,

\begin{equation}
\left|x_{0}| \approx |x_{0}(1)\right| (1 - 2Q^{2} S^{2}), \arg(x_{0}) = -2QS.
\label{eq:S5.6}
\end{equation}
The phase therefore departs from its linear value at $O(QS)\propto E^{2}$, whereas the amplitude departs only at $ O(Q^{2} S^{2}) \propto E^{4}$. With S = (3/8) $\xi$ in the quartic channel,

\begin{equation}
\delta\varphi = 2QS = (3/4) Q \xi, \Delta|x_{0}|/|x_{0}| = 2Q^{2} S^{2} = (9/32) Q^{2} \xi^{2}.
\label{eq:S5.7}
\end{equation}
At the convergence field $(\xi = \xi_{c} = 0.513/Q, Q = 9)$ these are $\delta\varphi = 0.385$ rad and $\Delta|x_{0}|/|x_{0}| = 0.074$, i.e. a phase departure of tens of degrees accompanied by only a few-percent amplitude roll-off. The experimental signature therefore depends on the observable: phase-resolved detection sees a departure $\propto (E/E_{\mathrm{conv}})^{2}$, amplitude detection one $\propto (E/E_{\mathrm{conv}})^{4}$. Two field ranges must be distinguished. At the onset, $\xi \ll \xi_{c}$, the phase departure is $\delta\varphi = (3/4) Q\xi$, so resolving the onset at $\xi \sim 10^{-3}$ calls for phase stability below $\sim 10^{-2}$ rad, within reach of electro-optic sampling \cite{hirori2011}. Well beyond the boundary, $\xi = 0.1\sim0.3$, the same expression gives $\delta\varphi \approx$ 0.7$\sim$2.0 rad and the effect is large. The small number quoted for the onset and the radian-scale number quoted deep inside the failed regime are therefore not in tension; what is "silent" is that none of this is announced by a jump, a hysteresis, or multistability. For $ \mathrm{LiNbO}_{3}, \xi_{c} = 0.057$.

\stepcounter{section}
\section*{S6. Experimental controls, and cascaded second-order contributions}
\subsection*{S6A. Experimental controls}
Four controls separate the convergence signal from competing effects.

{\bf{(i) Thermal versus coherent nonlinearity.}} A terahertz pulse deposits heat at rate $\alpha I$ per unit length; for LiNbO$_3$ at 7.6 THz $(\alpha \approx 10^{2}\sim10^{3}\,\mathrm{cm}^{-1})$ a 1 ps, 1 MV/cm pulse deposits of order 1 J/cm$^{3}$, a temperature rise of order 0.4 K per pulse. Fixing fluence while varying the repetition rate changes the thermal load by orders of magnitude while holding the coherent response per pulse fixed, so a shift that tracks the repetition rate is thermal and one that does not is coherent.

{\bf{(ii) Cascading.}} Varying the phase mismatch $\Delta k$ isolates the cascaded $\chi^{(2)}$ contribution (Sec. S6B), which scales as $ 1/\Delta k$.

{\bf{(iii) Phase matching.}} In a sample of length $L$ a mismatched process contributes $\propto \mathrm{sinc}(\Delta kL/2),$ which is non-monotonic in $L$; varying the propagation length distinguishes a phase-matching envelope from the intrinsic convergence roll-off.

{\bf{(iv) Domain reversal.}} Second-harmonic microscopy before and after the measurement, sensitive to the sign of $\chi^{(2)}$, excludes field-induced domain reversal.

\subsection*{S6B. Cascaded second-order contributions}
Cascaded $\chi^{(2)}:\chi^{(2)}$ processes, up-conversion to $ 2\omega$ followed by down-conversion to $\omega$ through a virtual intermediate state, generate an effective third-order response.

For LiNbO$_3$ the terahertz-band second-order coefficient $\chi^{(2)}$ is phonon-enhanced to $|\chi^{(2)}| > 1.58\times 10^{-6}\,\mathrm{m/V}$ at terahertz frequencies \cite{lu2021}. A cascade built from two such $\chi^{(2)}$ vertices is therefore non-negligible, and must be treated as an additional anharmonic contribution $\beta_{\mathrm{casc}}$ rather than dropped a priori. It does not, however, move the convergence radius: the cascade is itself a perturbative two-$\chi^{(2)}$-vertex process and only renormalizes the effective coefficient $\beta_{\mathrm{eff}}=\beta_{\mathrm{ionic}}+\beta_{\mathrm{casc}}$, while Eq.~(17) continues to bound the total response through $S$, which already contains the cascaded mode shift. Varying $\Delta k$ by angle or temperature tuning changes $\beta_{\mathrm{casc}}$ while leaving $\beta_{\mathrm{ionic}}$ fixed, allowing the two to be separated.

\vspace{3mm}
{\bf{Qualification}}

The statement that Eq.~(17) bounds the total response holds to the extent that the cascaded term feeds back into the measured frequency shift. Where the cascade contributes only to the polarization at $\omega$ without shifting the mode frequency, Eq.~(17) bounds the ionic part, and the two are separated by the $\Delta k$ scan of control (ii) in Sec. S6A.

\stepcounter{section}
\section*{S7. A second failure channel: k-photon resonance}
The boundary of Eqs. (9)-(17) is set by the fundamental response at the drive frequency $\omega$. An independent failure channel exists whenever a $k$-photon resonance $ k\omega = \omega_{\mathrm{res}}$ is met: the $k$-th-order susceptibility $\chi^{(k)}$ then has a pole, and the $k$-th-order polarization diverges even though the fundamental series is still convergent. Its field scale is

\begin{equation}
E(k\text{-photon}) \sim (\gamma_{\mathrm{res}} |\Delta_{k}|)^{1/2} / d_{k}, \Delta_{k} = k\omega - \omega_{\mathrm{res}},
\label{eq:S7.1}
\end{equation}
with $\gamma_{\mathrm{res}}$ the linewidth of the resonance and $ d_{k}$ an effective $k$-photon coupling. For a far-off-resonant intermediate state $(|\Delta_{k}| \gg \gamma_{\mathrm{res}})$ this channel does not compete with the fundamental boundary. For a near-resonant one it can tighten the bound. In LiNbO$_3$ the nearest higher infrared-active mode lies near 13 THz and the two-phonon continuum near 15 THz. This mode is of $E$ symmetry and does not couple to the driven $A_1$ mode through Eq.~(13). Within the window (5.8$\sim$8.8 THz) the relevant $k = 2$ condition (11.6$\sim$17.6 THz) can overlap these, but the two-phonon linewidth is several terahertz and the two-phonon coupling to the A1 mode is weak, so $E$($k$-photon) $\gg E_{\mathrm{conv}}$ over the range of Table I. The channel should be checked explicitly in materials with closer or more strongly coupled higher-lying resonances.

\vspace{3 mm}
{\bf{Spectral (Kubo) view of the same channel}}

A complementary view of the same channel is spectral. In the Kubo representation the $n$-th order term carries $n$ energy denominators, the $k$-th containing the $k$-photon detuning $\Delta_{k} = k\omega - \omega_{\mathrm{res}}$ \cite{kubo1957,orr1971}. A material that is transparent at $\omega$ can therefore fail through a $k$-photon resonance even when no mechanical resonance is nearby, and relaxation lowers the coherent bound \cite{kubo1957}. This is the same event seen from the susceptibility side rather than from the oscillator side, which is why the channel must be checked explicitly in materials with closer or more strongly coupled higher-lying resonances.

\stepcounter{section}
\section*{S8. Relation to previous limits on the perturbative regime}

\vspace{3 mm}
{\bf{Literature-search statement.}}

A systematic literature search found no previous derivation of a convergence-radius criterion for the driven anharmonic lattice mode. The closest works address (i) damage- or ionization-based limits on the perturbative regime \cite{zheltikov2019,khurgin2024}, and (ii) direct numerical summation of the polarization series in electronic epsilon-near-zero media, where no divergence was found \cite{reshef2017,reshef2019}. Neither derives a branch-point criterion of the kind given here.

\vspace{3 mm}
{\bf{Relation to Khurgin and Kinsey \cite{khurgin2024}}}

Those authors argue that a large $\chi^{\left( n \right)}$ does not by itself signal a non-perturbative response, since a series with large coefficients can still be summed. We agree, and address the distinct question they leave open, namely whether the series converges at all. Our criterion is fixed by the branch-point singularity of the exact response and is therefore independent of the magnitude of $\chi^{\left( n \right)}$ . A material may have very large low-order susceptibilities and still lie inside its convergence radius, or modest ones and already lie outside it. The two criteria are logically complementary, theirs constrains how large a truncated description has to be, ours constrains whether a truncated description exists at all.

\clearpage
\bibliography{refs}